\documentclass[preprint,a4paper,11pt,onecolumn]{elsarticle}

\usepackage{graphicx}
\usepackage{epsfig}
\usepackage{natbib}

\usepackage{url}
\usepackage{amsmath,amsfonts,bm,mathrsfs}
\usepackage{subfig}
\usepackage{multirow}
\usepackage{array}
\usepackage{multicol}
\usepackage{color}
\usepackage[normalem]{ulem}

\newtheorem{assumption}{Assumption}

\newtheorem{thm}{Theorem}
\newtheorem{lem}[thm]{Lemma}
\newtheorem{prop}[thm]{Proposition}
\newdefinition{rmk}{Remark}
\newproof{pf}{Proof}
\newproof{pot}{Proof of Theorem \ref{thm2}}

\journal{European Journal of Operational Research}

\begin{document}

\newcommand{\ue}{\bm{e}}
\newcommand{\ut}{\bm{\theta}}
\newcommand{\ub}{\bm{b}}
\newcommand{\uc}{\bm{c}}
\newcommand{\ua}{\bm{a}}

\newcommand{\vas}{\overline{v}}
\newcommand{\mycolor}[1]{\textcolor{red}{#1}}
\newcommand{\myst}[1]{\textcolor{red}{\st{#1}}}
\newcommand{\hT}{\hat{\Theta}}
\newcommand{\hatt}{\hat{\ut}}
\newcommand{\X}{\mathbb{X}}
\newcommand{\Y}{\mathbb{Y}}
\newcommand{\tX}{\tilde{\mathbb{X}}}
\newcommand{\R}{\mathbb{R}}
\newcommand{\C}{\mathbb{C}}
\newcommand{\ra}{\rightarrow}

\begin{frontmatter}



\title{Almost Budget Balanced Mechanisms with Scalar Bids For Allocation of a Divisible Good}


\author[a1]{D.~Thirumulanathan\corref{cor1}}
\ead{thirumulanathan@gmail.com}
\author[a2]{H.~Vinay}
\ead{vinayhebbatam@gmail.com}
\author[a3]{Srikrishna Bhashyam}
\ead{skrishna@ee.iitm.ac.in}
\author[a1]{Rajesh Sundaresan}
\ead{rajeshs@ece.iisc.ernet.in}

\cortext[cor1]{Corresponding author}

\address[a1]{Department of Electrical Communication Engineering, Indian Institute of Science, Bengaluru, India 560012}
\address[a2]{Audience Inc., 5th Floor (West wing), Umiya Business Bay, Tower 1, Cessna Business Park, Kadubeesanahalli, Varthur Hobli, Bengaluru, India 560087}
\address[a3]{Department of Electrical Engineering, IIT Madras, Chennai, India 600036}

\begin{abstract}
This paper is about allocation of an infinitely divisible good to several rational and strategic agents. The allocation is done by a social planner who has limited information because the agents' valuation functions are taken to be private information known only to the respective agents. We allow only a scalar signal, called a bid, from each agent to the social planner. \citet{YangHajek} and \citet{Johari09} proposed a scalar strategy Vickrey-Clarke-Groves (SSVCG) mechanism with efficient Nash equilibria. We consider a setting where the social planner desires minimal budget surplus. Example situations include fair sharing of Internet resources and auctioning of certain public goods where revenue maximization is not a consideration. Under the SSVCG framework, we propose a mechanism that is efficient and comes close to budget balance by returning much of the payments back to the agents in the form of rebates. We identify a design criterion for {\em almost budget balance}, impose feasibility and voluntary participation constraints, simplify the constraints, and arrive at a convex optimization problem to identify the parameters of the rebate functions. The convex optimization problem has a linear objective function and a continuum of linear constraints. We propose a solution method that involves a finite number of constraints, and identify the number of samples sufficient for a good approximation.
\end{abstract}


\end{frontmatter}


\section{Introduction}
\label{sec:introduction}

This paper is about allocation of an infinitely divisible good to several strategic agents. The social planner who does this allocation has limited information in the sense that the agents' valuation functions are taken to be private information known only to the respective agents. We allow only a scalar signal from the agents to the social planner, which we call a bid. This is the only means by which agents can provide information about their valuation functions to the social planner. We are interested in an {\em efficient} mechanism: the allocation should maximize the sum of valuations of the agents. Under these constraints, we study mechanisms that come close to budget balance. Example situations described next, include fair sharing of Internet resources, disbursal of funds by a parent department, and auctioning of certain public goods, where revenue maximization is not a consideration. 

{\em Example 1.} A communication channel with total capacity $C$ is to be shared among several rational and strategic agents. This channel can be allocated via a randomized allocation rule, and is thus an infinitely divisible resource. If an agent gets a long term average throughput of $a_i$, the agent's valuation is $v_i(a_i)$, where $v_i : [0,C] \rightarrow\mathbb{R}_+$ is increasing, concave, and known only to the agent. Naturally, $\sum_i a_i \leq C$. The agents wish to share the resources among themselves without money transferred to an external agent. Suppose that the agents agree to communicate with an external coordinator who attempts to maximize the sum of valuations. The signal space complexity to signal the valuation functions to the coordinator is prohibitive, particularly when the agents are geographically separated, because the functions can be arbitrary within the infinite-dimensional class of increasing concave functions. To model this communication constraint, we assume that the agents can send only a scalar signal. In this example, the coordinator is the social planner who desires efficient allocation without an interest in maximizing revenue. The scalar signals are viewed as bids.

{\em Example 2.} A parent organization has to disburse available funds (assumed divisible) among several of its departments. Each department has a certain valuation function $v_i$ for the allocation, is strategic, and the parent department desires to allocate efficiently while retaining only a minimal balance, if at all, based on limited information that the departments provide. Consider the extremely limited information setting of a scalar signal. The parent department is the social planner, the scalar signals are the bids, and the parent department desires an efficient distribution and no surplus.

The Vickrey-Clarke-Groves (VCG) mechanism (\citet{Vickrey,Clarke,Groves}) achieves efficient allocation, but only when the signal space is sufficiently complex to describe entire valuation functions. In the VCG mechanism, the social planner requests agents to submit their valuation functions. The social planner then allocates to maximize the sum of the submitted valuation functions and determines the agents' payments.

Motivated by the communication network context but with nonstrategic agents, \citet{Kelly97} proposed a mechanism that involved only scalar bids. Under the Kelly mechanism, the social planner first collects scalar bids from the agents. Then the social planner allocates the good in proportion to the bids, and collects payments equal to the bids. The price per unit, or the market clearing price, is the sum of the bids divided by the quantity of the good. Every agent sees the same market clearing price. This distributed solution was shown to be efficient under certain conditions, but the agents should be price-taking or nonstrategic. If the agents are strategic, there is an efficiency loss of up to, but not more than, $25\%$ \citep{Johari2004}.

The VCG mechanism payments involve prices per unit good that can differ across the agents. This is not the case in the Kelly mechanism. In order to reduce the efficiency loss in strategic settings with scalar bids, \citet{YangHajek} and \citet{Johari09} brought the feature of price differentiation across agents (a feature of the VCG mechanism) to the Kelly mechanism. The resulting mechanism, a scalar strategy VCG mechanism\footnote{For some examples of mechanism design with restricted signaling, see \citet{reichelstein1988game} (minimal strategy space dimension for fully efficient Nash equilibria), \citet{semret1999market} (two-dimensional bids for each resource), \citet{jain2010efficient} (two-dimensional bids on bundles of resources), \citet{blumrosen2007auctions} (number of bits needed for signaling the bid). Our focus however is on the one-dimensional signaling.} (SSVCG), was shown to have efficient Nash equilibria.

All the above mechanisms typically result in a budget surplus (sum of payments from agents is positive). In this paper, our ideal is to achieve {\em budget balance}, or zero budget surplus. However, simultaneously achieving efficiency and budget balance in a strategy-proof mechanism is, in general, not possible (due to the Green-Laffont theorem \citep{GreenLaffont1977}; see footnote \ref{footnoteGFT}).

In the VCG setting, where there is no constraint on signaling, various {\em almost budget balance} notions and associated mechanisms were proposed. Almost budget balance is achieved by redistributing the payments among the agents in the form of rebates. \citet{Guo} and \citet{Moulin} studied rebate design in the case of discrete goods. \citet{GujNar09,GujNar11} studied rebate design for the allocation of $m$ heterogeneous discrete goods among $n$ agents. \citet{Chorppath} studied rebate design in the divisible goods setting.

A big advantage with the VCG setting is that the social planner comes to know the true valuation functions. Voluntary participation of agents, i.e., agents being better off by participating in the mechanism, is easily verified. Furthermore, knowledge of the valuation functions could be exploited in defining a criterion for almost budget balance, as is done in \citet{Moulin} and \citet{Chorppath}. The extension of the almost budget balance notion to the SSVCG setting, however, is not straightforward. We cannot assume that the valuation functions are available because agents supply only a scalar bid. We thus relax our objective to that of achieving Nash equilibrium instead of achieving the DSIC (Dominant Strategy Incentive Compatibility) property.

In this paper, we consider the SSVCG setting that allows the agents to send only a scalar bid. We (1) propose a notion of {\em almost budget balance} appropriate for the SSVCG setting, and (2) design an optimal mechanism as per the proposed notion of almost budget balance.

\citet{kakhbod-2012} designed a mechanism to achieve an efficient Nash equilibrium with no budget surplus, but considered a setting where the agents signal a two-dimensional bid to the social planner. Moreover, their mechanism may not be feasible when the signals of the agents are not at Nash equilibrium. \citet{sinha-2013} modified this mechanism to have feasibility even under off-equilibrium situations, but required agents to signal a four-dimensional bid to achieve strong budget balance at equilibrium. We are not aware of any mechanism that achieves an efficient Nash equilibrium with strong budget balance using only scalar bids.

There are several design choices that we will make in arriving at a criterion for almost budget balance in the SSVCG setting. Considerations of tractability and significant reduction in surplus will guide our design decisions. For example, we restrict attention to the so-called linear rebates. This is mainly because it makes the optimization problem analytically tractable. An additional reason for the choice of linear rebates is that they are known to be optimal in the homogeneous discrete goods setting of \citet{Moulin} and \citet{Guo}. The best justification however is the significant reduction in the surplus seen in our simulation results.

The coefficients of the linear rebate functions will be determined by a solution to a convex optimization problem. Specifically, we need to solve an {\em uncertain convex program} (UCP) \citep{ref:Calafiore03} involving a linear objective function and a continuum of linear constraints. We propose a solution method that involves a finite number of constraints, and provide guarantees on the number of samples needed for a good approximation. We first prove that, under some sufficient conditions, the solutions of a general UCP and its corresponding relaxed UCP are close. We then prove that the specific linear rebate UCP satisfies these sufficient conditions.

The rest of this paper is organized as follows. In Section \ref{sec:problem_setup}, we discuss the problem setting and the SSVCG mechanism. In Section \ref{sec:design-considerations}, we discuss design choices for almost budget balance and rebate functions, our design decisions, and formulate an optimization problem. In Section \ref{sec:simplify}, we make crucial reductions that ensure that our proposal can be implemented. The resulting optimization problem is a UCP. In Section \ref{sec:UCP}, we study a general UCP and formulate a sufficient condition for an approximate solution via sampling of constraints. In Section \ref{sec:apply-UCP}, we apply the solution of Section \ref{sec:UCP} to the UCP for almost budget balance. In Section \ref{sec:discussion}, we summarize our results, discuss alternative choices, and suggest possible extensions. Some simulation results demonstrate the usefulness of our approach.

\section{The setting}
\label{sec:problem_setup}

\subsection{SSVCG Mechanism}
\label{subsec:ssvcg}
A social planner needs to allocate a unit divisible resource among $n$ intelligent, rational, and strategic agents. Agent $i$ has a {\em valuation function} $v_i : [0,1] \ra \R_+$ privately known only to herself. The interpretation is that if $a_i \in [0,1]$ is the fraction of the good allocated to agent $i$, her valuation is $v_i(a_i)$. The social planner's goal is to solve the following problem:
\begin{equation} \label{SYSTEM}
	\max_{\{a_i\}} \sum_{i=1}^n v_i(a_i)\hspace*{.3in}\text{subject to}\hspace*{.3in}\sum_{i=1}^n a_i \leq 1,\text{ and }a_i \geq 0 \ \forall i.
\end{equation}
The social planner, however, does not know the valuation profile $v_1, \ldots, v_n$. To get some indication of these from the agents, the social planner chooses the following mechanism with one-dimensional signals from the agents. The social planner announces, a priori, a scalar-parametrized {\em surrogate valuation function} set $\mathcal{V} = \{\vas(\cdot, \theta), \theta \in [0, \infty)\}$.  The function $\vas(\cdot, 0)$ is taken to be the zero function. An agent $i$ is asked to bid $b_i \in [0,\infty)$, which is taken to be a signal of that agent's desired surrogate valuation function $\vas(\cdot,b_i)$. All agents bid simultaneously. The bid profile is denoted $\ub = (b_1, \ldots, b_n)$. If $\ub$ is the all-zero vector, the social planner allocates nothing. Otherwise, the social planner allocates the divisible good by solving the following problem which is naturally analogous to (\ref{SYSTEM}) but arising from the signaled surrogate valuation functions:
\begin{equation} \label{Soc_plan_prob_a}
	\max_{\{a_i\}} \sum_{i=1}^n \vas(a_i,b_i)\hspace*{.3in}\text{subject to}\hspace*{.3in}\sum_{i=1}^n a_i \leq 1,\text{ and }a_i \geq 0 \ \forall i.
\end{equation}
A payment $p_i(\ub)$ is then imposed on agent $i$. This payment is given by
\begin{equation}
  \label{eqn:VCG-payment}
  p_i(\ub) = - \sum_{j \ne i} \vas(a_j^*,b_j) + \sum_{j\ne i} \vas(a_{-i,j}^*,b_j) - r_i(\ub_{-i}),
\end{equation}
where the terms $a_j^*$, $a_{-i,j}^*$, and $r_i$ are as explained next. The term $a_j^*$ denotes the $j^{th}$ coordinate of the optimal solution to the social planner problem in \eqref{Soc_plan_prob_a}. Its dependence on $\ub$ is understood and suppressed. Similarly, $ a_{-i,j}^* $ is the $ j^{th} $ component of the optimal allocation when agent $i$ is not participating in the mechanism. Its dependence on $ \ub_{-i} $, the bids of all agents other than agent $i$, is once again understood and suppressed. The function $r_i$ is arbitrary and has as its argument the bids of all agents other than $i$. Agent $i$'s resulting {\em quasi-linear} utility is $v_i(a_i^*(\ub)) - p_i(\ub)$.

\begin{figure}
\begin{center}
\includegraphics[scale=1]{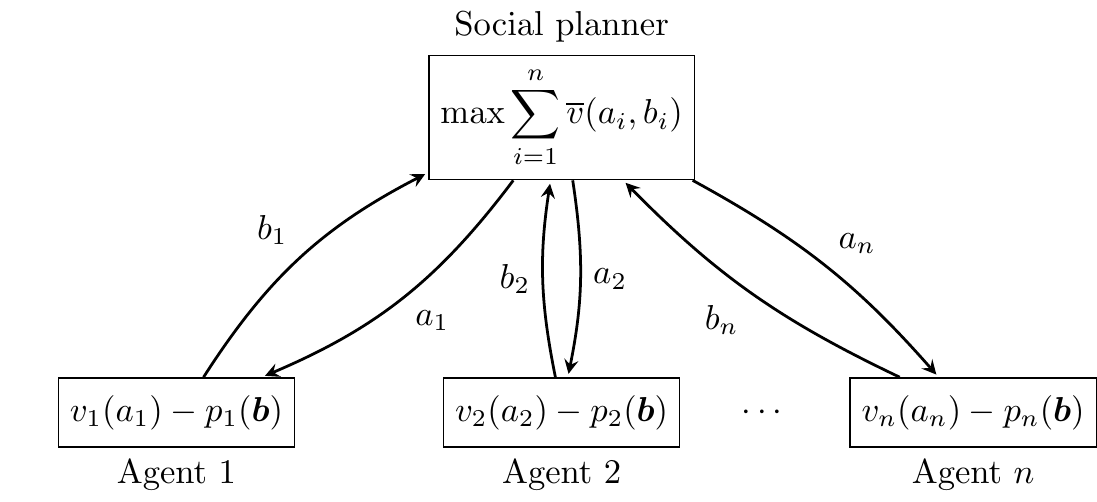}
\caption{Schematic representation of the SSVCG mechanism}
\label{schematic}
\end{center}
\end{figure}

With the above specifications, we have a simultaneous action game (with incomplete information) among the $n$ agents. A schematic illustrating the problem solved by the social planner, the utilities of the agents, and the exchange of information is shown in Figure \ref{schematic}. Since each agent's strategy is to choose a one-dimensional or scalar bid, and since the payments are inspired by the VCG mechanism, this mechanism is called the scalar-strategy VCG (SSVCG) mechanism\footnote{Scalar-strategy Groves mechanism is perhaps more appropriate. We will stick to the terminology of \citet{Johari09}.}. The first two terms of the right-hand side of (\ref{eqn:VCG-payment}) constitute the payment of agent $i$ in Clarke's pivotal mechanism, and the last term $r_i(\ub_{-i})$ may be viewed as a rebate given to agent $i$.

The VCG mechanism satisfies DSIC; it is in the best interest of each agent to signal her valuation function in its entirety. In our SSVCG setting, however, the signal dimension is greatly reduced because only a scalar bid is permitted. Incentive compatibility is not possible in general\footnote{\label{foot1}Incentive compatibility is possible in some special settings. Consider the following restricted VCG setting where, for simplicity, it is common knowledge that $v_i \in \mathcal{V}$ for $i = 1, 2, \ldots, n$. Then, for each $i$, there is a $\theta_i$ such that $v_i(\cdot) = \vas(\cdot, \theta_i)$. The private information held by agent $i$ is the scalar $\theta_i$, and it can be seen that $\ub = \ut = (\theta_1, \theta_2, \ldots, \theta_n)$ is an equilibrium in dominant strategies.}, and we shall settle for a Nash equilibrium. The following assumptions suffice to guarantee the existence of, not just a Nash equilibrium, but an {\em efficient Nash equilibrium}\footnote{A Nash equilibrium is efficient if it yields an allocation that solves \eqref{SYSTEM}.} \citep[Cor. 1]{Johari09}, and furthermore, to assert that every Nash equilibrium is efficient \citep[Prop. 2]{Johari09}.
\begin{assumption}
\label{ass_efficient_NE}
\begin{itemize}
  \item[(a)] For each $i$, $v_i$ is concave, strictly increasing, and continuously differentiable\footnote{The derivatives at the end-points are one-sided, with $\infty$ as a possible value at 0.} on $[0,1]$, with $v_i(0) = 0$. Moreover, at least two agents have infinite marginal valuations at 0, that is, there exist two agents $i,j$  with $i \ne j$ such that $v_i'(0)=v_j'(0)=\infty$.
  \item[(b)] For every $\theta > 0$, the function $\vas(\cdot,\theta)$ is strictly concave, strictly increasing, and continuously differentiable over $[0,1]$, with $\vas(0,\theta) = 0$ for all $\theta \geq 0$. Furthermore, for any $\theta > 0$, the derivative with respect to the first argument satisfies $\vas'(0,\theta) = \infty$.
  \item[(c)] For every $\gamma>0$ and $a > 0$, there exists $\theta > 0$ such that $\vas'(a,\theta) = \gamma$.
\end{itemize}
\end{assumption}
Specifically, let us restrict the surrogate valuation function to be of the form $\vas(a,\theta)=\theta U(a)$, where $U:[0,1]\rightarrow\mathbb{R}$ satisfies the following assumptions.
\begin{assumption}
\label{ass_surrogate}
$U$ is strictly concave, strictly increasing, and a continuously differentiable function over $[0,1]$ with $U(0)=0$ and $U'(0)=\infty$.
\end{assumption}
It is easy to verify that the surrogate valuation functions of the above form satisfy Assumptions \ref{ass_efficient_NE}(b) and \ref{ass_efficient_NE}(c). The reason for this restriction is technical, and will be clear in Section \ref{subsec:simplifyFW}.

\subsection{Almost budget balanced SSVCG mechanism}
\label{subsec:abb}
When $r_i$ in (\ref{eqn:VCG-payment}) is identically zero, Clarke's pivotal payment rule may result in a net budget surplus (sum of payments) at the social planner. In this paper, however, we are interested in scenarios where the social planner wants efficient allocation, but desires zero budget surplus. Zero budget surplus, also called {\em budget balance}, is unattainable in general\footnote{\label{footnoteGFT}The Green-Laffont theorem \citep{GreenLaffont1977} says that, when the set of valuation functions are sufficiently rich, there is no quasi-linear mechanism that simultaneously satisfies DSIC, allocative efficiency, and budget balance properties. Take the restricted VCG setting in footnote \ref{foot1}. The Green-Laffont theorem is applicable to this setting, and since the mechanism is DSIC and allocatively efficient, it follows that budget balance is impossible.}. Our objective is to achieve {\em almost budget balance}, a notion we will formalize in this section.

Two properties are desired for these mechanisms after payments are collected and rebates are redistributed as in (\ref{eqn:VCG-payment}). They are:
\begin{enumerate}
\item[({\bf F})] \textbf{Feasibility or weak budget balance}: The mechanism should not be subsidized by an external money source. This imposes the constraint that, for each $\ub$, we should have
$$\sum_{i=1}^{n} p_i(\ub) \geq 0, $$
which, using (\ref{eqn:VCG-payment}), is seen to be equivalent to
\begin{eqnarray}
  \sum_{i=1}^{n} r_i(\ub_{-i}) & \leq & \sum_{i=1}^{n}\left[- \sum_{j\ne i} \vas(a_j^*,b_j)\right. + \left.\sum_{j\ne i} \vas(a_{-i,j}^*,b_j)\right] \nonumber \\
  & = & - (n-1) \sum_{j=1}^n \vas(a_j^*,b_j) + \sum_{i=1}^{n} \sum_{j \ne i} \vas(a_{-i,j}^*,b_j) \nonumber \\
  & =: & p_{S}(\ub), \label{eqn:p-S}
\end{eqnarray}
where $ p_{S}(\ub) $ is the total surplus under Clarke's pivotal payment rule.

\item[({\bf VP})] \textbf{Voluntary participation}: Agents should be better-off (in the sense of not being strictly worse-off) by participating in the mechanism. We take the payoff (utility) for not participating in the mechanism to be 0. (VP) then imposes the constraint that, for each $\ub$, we should have
\begin{equation}
  \label{eqn:VP-S1}
  v_i(a_i^*) - p_i(\ub) \geq 0, ~\forall i,
\end{equation}
which, using (\ref{eqn:VCG-payment}) once again, is equivalent to
\begin{eqnarray}
  \label{eqn:VP-S}
  r_i(\ub_{-i})
  & \geq & -v_i(a_i^*)-\sum_{j \ne i}\vas(a_j^*,b_j)+\sum_{j \ne i}\vas(a_{-i,j}^*,b_j), ~\forall i \nonumber \\
  & =: & q_i(\ub), \label{eqn:ni}
\end{eqnarray}
\end{enumerate}
where $q_i(\ub)$ is the negative of the quasi-linear utility of agent $i$ under Clarke's payment rule.

An issue now arises. While the payments $p_i(\ub)$ do not depend explicitly on the true valuation function, which is unknown to the social planner, the condition for (VP) does. This can be seen  in (\ref{eqn:VP-S}) by observing that $q_i(\ub)$ depends on $v_i$.

When $r_i(\cdot) \equiv 0$, $i = 1, 2, \ldots, n$, Clarke's pivotal mechanism satisfies both (F) and (VP). Are there other mechanisms with nontrivial rebate functions that satisfy (F) and (VP)? We shall answer in the affirmative in Section \ref{sec:apply-UCP}, and we shall see how the issue of dependence of (VP) on $v_i$, which the social planner does not know, is addressed in Section \ref{subsubsec:simplifyVP}.

\section{Design considerations leading to an optimization problem}
\label{sec:design-considerations}

\subsection{Deterministic and anonymous rebates}
\label{subsec:d&a}

For a given set of bids, we require that the rebates be {\em deterministic}: the mechanism does not employ randomness. Additionally, we require that the rebates be {\em anonymous}: two agents with identical bids should receive identical rebates. The information available to the social planner on the valuation functions is symmetric across agents. Indeed, all that the social planner knows is that the valuation functions satisfy Assumption \ref{ass_efficient_NE}(a). This information is symmetric to permutation of agent labels. After the bids are sent to the social planner, two agents with identical bids are indistinguishable, and so, we require that the mechanism give them identical rebates.

To ensure deterministic and anonymous rebates, we restrict attention to rebates of the following form. For a bid profile $\ub$, let $\ub_{[j]}$ be the $j^{\text{th}}$ largest entry of $\ub$. Similarly, for $\ub_{-i}$, let $(\ub_{-i})_{[j]}$ be the $j^{\text{th}}$ largest entry of $\ub_{-i}$. The rebate functions are taken to be of the form\footnote{\label{footnoteConverse}This choice is motivated by the following observation. Consider the restricted VCG setting of footnote \ref{foot1} where a converse statement holds: if
\begin{equation}
  \label{eqn:holmstromaya-namaha}
  \sup_{(a,\theta) \in [0,1]^2} \left| \frac{\partial \vas(a, \theta)}{\partial \theta} \right| < \infty,
\end{equation}
then any DSIC mechanism with a rebate function that is deterministic and anonymous must have the form (\ref{eqn:det-anon}). This converse was informally stated by \citet{Cavallo06}, and formally proved in \citep[Lem. 2]{Guo}. The proof relies on a result of \citet{Holmstrom79} that shows that any DSIC mechanism must be of the Groves class if the family of valuation functions is `smoothly connected'. The latter property holds for our single-parameter family of surrogate valuation functions when (\ref{eqn:holmstromaya-namaha}) holds.}
\begin{equation}
  \label{eqn:det-anon}
  r_{i}(\ub_{-i}) = g((\ub_{-i})_{[1]},(\ub_{-i})_{[2]},\ldots,(\ub_{-i})_{[n-1]}).
\end{equation}
In following subsections, we propose optimality criteria for designing rebates.

\subsection{Design for the worst case}
\label{subsec:des-for-worst-case}

Suppose that there are $m$ discrete and identical goods, and each agent is allocated at most one good. The valuation function of agent $i$ may be taken to be $\theta_i a_i$, where $a_i \in \{0, 1\}$. The private information $\theta_i$ is then interpreted as the value of the good to agent $i$. Clearly, this is a setting where, with $U(a_i)=a_i$, the proposed mechanism is just the VCG mechanism and achieves DSIC. We may therefore take the bids to be $b_i = \theta_i$ for each agent $i$. For this setting, \citet{Moulin} defined {\em almost budget balance} in terms of the {\em worst-case} ratio of the sum of payments to the sum of valuations. Specifically, Moulin's proposal is to design rebates to minimize
\begin{equation}
  \label{eqn:Moulin-wc-abb}
  \sup_{\ut}\left[\frac{p_S(\ut) - \sum_{i=1}^n r_i(\ut_{-i})}{\sigma(\ut)}\right],
\end{equation}
where $\sigma(\ut) = \sum_{i=1}^n \vas(a_i^*,\theta_i)$ is the optimal social welfare, subject to:
\begin{align}
    \label{eqn:F}
	\text{(F)}  \quad & \sum_{i=1}^n r_i(\ut_{-i}) \leq p_S(\ut) \quad \forall \ut,\\
    \label{eqn:VP}
	\text{(VP)} \quad & r_i(\ut_{-i})  \geq q_i(\ut) \quad \forall i, \quad \forall \ut.
\end{align}
See (\ref{eqn:p-S}) and (\ref{eqn:ni}) for definitions of $p_S$ and $q_i$, respectively. \citet{Guo} considered an alternate proposal to minimize
\begin{equation}
  \label{eqn:guo-conitzer-proposal}
  \sup_{\ut}
   \left[ 1 - \frac{\sum_{i=1}^n r_i(\ut_{-i})}{p_S(\ut)} \right]
\end{equation}
subject to the same (F) and (VP) constraints.

It turns out that in the above example of an auction of $m$ identical discrete goods, considered both by \citet{Moulin} and \citet{Guo}, the two proposals yield the same optimal rebates and objective function values. In general, however, the two proposals yield different solutions. Indeed, they yield different solutions for the auction of $m$ identical discrete goods if, for example, the (VP) constraint alone is relaxed. The Guo and Conitzer proposal focuses only on the fraction of payment that is retained as net surplus without regard to the absolute value of the payment amounts. A rewriting of Moulin's objective (\ref{eqn:Moulin-wc-abb}) as
\[
   \sup_{\ut}
   \left[ \frac{p_S(\ut)}{\sigma(\ut)} \cdot \left( 1 - \frac{\sum_{i=1}^n r_i(\ut_{-i})}{p_S(\ut)} \right) \right]
\]
clearly shows that the fraction of retained surplus, the quantity considered by Guo and Conitzer and enclosed within parentheses above, is weighted by a factor that takes into account the size of the payments $p_S(\ut)$ relative to the optimal social welfare $\sigma(\ut)$. If the Guo and Conitzer proposal attains its worst-case at a profile where the net Clarke surplus is small relative to the optimal social welfare, it is de-emphasized by the Moulin proposal. The Moulin proposal, therefore, focuses more on reducing the surplus in settings where the net surplus is high relative to the optimal social welfare. We therefore adopt Moulin's proposal of minimizing (\ref{eqn:Moulin-wc-abb}).

\citet{Chorppath} studied exactly this proposal in the divisible goods setting, but in the simpler restricted VCG setting of footnote \ref{foot1}. While the Moulin proposal is defensible for that setting, it has the drawback in our SSVCG setting that $\sigma(\ut)$ is not known to the social planner. Just as we chose a surrogate valuation function to identify the allocations and payments, we choose a surrogate social welfare function
\begin{equation}
\sigma_S(\ub) := \sum_{i=1}^n \vas(a_i^*,b_i)
\label{eqn:sigma-S}
\end{equation}
in place of $\sigma(\ub)$. We therefore propose to minimize
\begin{alignat}{2}
\label{eqn:Moulin-wc-abb-ssvcg}& \sup_{\ub} \left[\frac{p_S(\ub) - \sum_{i=1}^n r_i(\ub_{-i})}{\sigma_S(\ub)}\right]\\
\label{eqn:F-ssvcg} \text{subject to} \hspace*{.5in} & \text{(F)}  \quad \quad \sum_{i=1}^n r_i(\ub_{-i}) \leq p_S(\ub) \quad \forall \ub,\\
\label{eqn:VP-ssvcg}& \text{(VP)} \quad \, r_i(\ub_{-i})  \geq q_i(\ub) \quad \forall \ub, \quad \forall i,
\end{alignat}
where (\ref{eqn:F-ssvcg}) is the same as (\ref{eqn:p-S}), and (\ref{eqn:VP-ssvcg}) is the same as (\ref{eqn:VP-S}). This choice puts us in the optimization framework of \citet{Chorppath}, except that we have not yet shown how to resolve the issue of dependence of (\ref{eqn:VP-ssvcg}) on the private information $v_i$.

We now highlight two important differences between our work and that of \citet{YangHajek}. (1) For given valuation functions, they bound the payments and revenues of the SSVCG mechanism, and remark that there are suitable surrogate functions that can drive the revenue to zero. For a given surrogate function, however, by merely scaling the (true) valuation functions, the revenue can be made arbitrarily large. Thus their analysis does not address the worst case setting while ours does. (2) Neither \citet{YangHajek} nor \citet{Johari09} explicitly discuss or impose the VP constraint, while we do. But, as we will point out after Lemma \ref{lem:simplifyVP}, the SSVCG mechanism with Clarke's pivotal payment rule does indeed satisfy the VP constraint at Nash equilibrium if there is a bid, say $0$, that signals withdrawal from the mechanism.

\subsection{Linear rebates}
\label{subsec:linear-rebates}

In the \citet{Moulin} and \citet{Guo} settings, which is that of worst-case optimal rebates for the auction of discrete identical goods, linear rebate functions of the form
\begin{equation}\label{eqn:lin_rebate_unordered}
  r_i(\ub_{-i}) = c_0 + c_1 (\ub_{-i})_{[1]} + \ldots + c_{n-1}(\ub_{-i})_{[n-1]}
\end{equation}
were shown to be optimal. We too restrict attention to linear rebates of this form. Optimality or otherwise of linear rebates for the divisible good case is still unexplored. Linear rebates enable analytical tractability as we will see in later sections. In Section \ref{sec:discussion}, we present some numerical results that justify to some extent the use of linear rebates.

\subsection{Restriction to the closure of realizable signals}
\label{subsec:restriction-closure}

The performance metric (\ref{eqn:Moulin-wc-abb-ssvcg}) has a supremum over $\ub$ subject to constraints (\ref{eqn:F-ssvcg}) and (\ref{eqn:VP-ssvcg}) which are also parametrized by $\ub$. The supremum and the constraints ought to reflect only those $\ub$ that are {\em realizable}, i.e., those $\ub$ that are Nash equilibria for some valuation function profiles $v_1 \ldots, v_n$. We now identify this set of realizable points.

\begin{lem}
\label{NE is Full Rn} Let the surrogate valuation function satisfy Assumptions \ref{ass_efficient_NE}(b) and \ref{ass_efficient_NE}(c). Then, for any $\ut \in (0,\infty)^n$, there exist valuation function profiles $v_1, \ldots, v_n$ satisfying Assumption \ref{ass_efficient_NE}(a) such that $\ub = \ut$ is a Nash equilibrium.
\end{lem}

\begin{pf}
Proof is available in the online appendix.\qed
\end{pf}

The form of the surrogate valuation function implies some regularity on $p_S$ and $\sigma_S$.

\begin{lem}
\label{lem:sigmasps-lipschitz}
With $\vas(a,\theta) = \theta U(a)$, where $U$ satisfies Assumption \ref{ass_surrogate}, the mappings $\ub \mapsto \sigma_S(\ub)$ and $\ub \mapsto p_S(\ub)$ are Lipschitz continuous.
\end{lem}

\begin{pf} For $\sigma_S$, see Thm.~4 of \citet{Chorppath}. For $p_S$, see Lem.~1 of \citep{Chorppath}. The proofs are reproduced in the online appendix for completeness.\qed
\end{pf}

Lemmas \ref{NE is Full Rn} and \ref{lem:sigmasps-lipschitz}, and the fact that our choice of linear rebates is continuous in $\ub$, allow us to run $\ub$ over the closure of the set of realizable (or Nash equilibrium) bids. Since this closure is $\mathbb{R}_+^n$, $\ub$ runs over all elements in $\mathbb{R}_+^n$, both in the supremum and in the (F) and (VP) constraints.

\subsection{Ordering and the optimization problem}
\label{subsec:ordering}

Observe that the worst-case optimality criterion (\ref{eqn:Moulin-wc-abb-ssvcg}) depends only on the ordered bids. Without loss of generality, we henceforth assume that agent $i$ is the agent with the $i^{\text{th}}$ highest bid. Bids then come from the set $\hT = \{ \ub \in \mathbb{R}_+^n ~|~ b_1 \geq b_2 \geq \ldots \geq b_n \geq 0 \}$, and the $i^{\text{th}}$ agent's rebate is
\begin{equation}\label{eqn:lin_rebate}
  r_i(\ub_{-i}) = c_0 + c_1b_1 + \ldots + c_{i-1}b_{i-1} + c_{i}b_{i+1} + \ldots + c_{n-1}b_n.
\end{equation}

Henceforth, when we refer to the optimization problem in (\ref{eqn:Moulin-wc-abb-ssvcg}) subject to the (F) and (VP) constraints in (\ref{eqn:F-ssvcg}) and (\ref{eqn:VP-ssvcg}), we replace the parameter $\ub$ by $\ut$, and $\forall\ub$ by $\forall\ut\in\hat{\Theta}$. Let us also define $\uc = (c_0, \ldots, c_{n-1})$.

The optimization problem to design the best linear rebate functions, after substitution of \eqref{eqn:lin_rebate} in (F) of (\ref{eqn:F-ssvcg}) and in (VP) of (\ref{eqn:VP-ssvcg}), is now:
\begin{alignat}{2}
\label{eqn:ordered-opt-prob} & \min_{\uc}  \sup_{\ut \in \hT} \ \left[\frac{p_S(\ut) - \sum_{i=1}^n r_i(\ut_{-i})}{\sigma_S(\ut)}\right]\\
\label{eqn:F-ordered}\text{subject to }\hspace*{.2in}& \text{(F)}  \quad\, \ nc_{0}+\sum_{i=1}^{n-1}c_{i}(i\theta_{i+1}+(n-i)\theta_{i})\le p_S(\ut), \ \forall\ut \in \hT\\
\label{eqn:VP-ordered}&\text{(VP)} \quad \, c_{0}+\sum_{j=1}^{i-1}c_{j}\theta_{j}+\sum_{j=i}^{n-1}c_{j}\theta_{j+1}\ge q_i(\ut), \ \forall \ut \in \hT, \forall i.
\end{alignat}

\section{Simplification of constraints and a reformulation}
\label{sec:simplify}

We now free up the optimization problem in (\ref{eqn:ordered-opt-prob}) from its dependence on the true valuation functions in the (VP) constraint. We also justify the restriction of $\ut$ to a compact subset of $\hT$ and arrive at a reformulation of the above optimization problem as a generalized linear program.

\subsection{Simplification of the (VP) constraints}
\label{subsubsec:simplifyVP}

As observed earlier, the (VP) constraint in (\ref{eqn:VP-ordered}) requires, through $ q_i(\ut) $, knowledge of the true valuation functions. The following lemma is a significant step in freeing up the constraint from the knowledge of true valuation functions, and assures us that the optimization problem is well-posed. Incidentally, this will also establish that the SSVCG mechanism with Clarke's pivotal payment rule satisfies the VP constraint at Nash equilibrium.

\begin{lem}
\label{lem:simplifyVP}
Suppose that the true valuations satisfy Assumption \ref{ass_efficient_NE}(a) and that the surrogate valuation function is $\vas(a,\theta) = \theta U(a)$, with $U$ satisfying Assumption \ref{ass_surrogate}.
\begin{itemize}
  \item[(1)] The constraints (F) and (VP), (\ref{eqn:F-ordered}) and (\ref{eqn:VP-ordered}), imply that $c_0 = c_1 = 0$.

  \item[(2)] Let $c_0 = c_1 = 0$. Then, the (VP) constraint is equivalent to
  \begin{equation}
    \label{eqn:simplify-VP}
    \sum_{i=2}^k c_i \geq 0, \mbox{ for } k = 2, 3, \ldots, n-1.
  \end{equation}
\end{itemize}
\end{lem}

\begin{pf}
 Proof is available in the online appendix.\qed
\end{pf}

Thanks to Lemma \ref{lem:simplifyVP}, the optimization problem is now given by:
\begin{alignat}{2}
\label{eqn:VP-simplified-opt-problem} & \min_{\uc} \sup_{\ut \in \hT} \ \left[\frac{p_S(\ut) - \sum_{i=1}^n r_i(\ut_{-i})}{\sigma_S(\ut)}\right]\\
\text{subject to }\hspace*{.5in} & \text{(F)} \quad \quad \ \sum_{i=2}^{n-1}c_{i}(i\theta_{i+1}+(n-i)\theta_{i})\le p_S(\ut), \ \forall\ut \in \hT \nonumber\\
  & \text{(VP)} \quad \,\,\,\, \sum_{i=2}^k c_i \geq 0, \ k = 2, 3, \ldots, n-1.\nonumber
\end{alignat}

\subsection{Reformulation as a generalized linear program}
\label{subsec:LP}

As in \citet{Chorppath}, the min-max problem (\ref{eqn:VP-simplified-opt-problem}) can be turned into a generalized linear program (LP) by introducing an auxiliary variable $t$:
\begin{alignat*}{2}
&\min_{\uc,t} \ t\\
\text{subject to }\hspace*{.3in}
		&\text{(F)}  \quad \sum_{i=2}^{n-1}c_{i}(i\theta_{i+1}+(n-i)\theta_{i})\le p_S(\ut), \ \forall\ut \in \hT \\
		&\text{(VP)} \, \sum_{i=2}^k c_i \geq 0, \ k = 2, 3, \ldots, n-1, \\
		&\text{(W)}  \,\,\,\, \sum_{i=2}^{n-1}c_{i}(i\theta_{i+1}+(n-i)\theta_{i}) + t \sigma_S(\ut) \ge p_S(\ut), \forall \ut \in \hT.
\end{alignat*}
(W) captures the constraint associated with the worst-case objective.

We say ``generalized'' because the above LP has a {\em continuum} of linear constraints parametrized by $\ut \in \hT$. The constraint set on $\uc$ and $t$ is convex, because it is an intersection of a family of half-plane constraints. While there appears to be no direct way to solve this problem, further simplification of the constraints is possible. We pursue this in the next subsection.

\subsection{Simplification of (F) and (W)}
\label{subsec:simplifyFW}

In \ref{app:monotonicity} we show two properties -- monotonicity and scaling -- of the VCG payments. We shall now exploit them to simplify (F) and (W).

Observe that the left-hand side of (F) does not depend on $\theta_1$. From Proposition \ref{prop:monotone-scaling}(a), we have that $p_S(\ut)$ is monotonically increasing in $\theta_1$ for a fixed $\ut_{-1}$. It follows that the right-hand side is smallest (and the constraint is tightest) when $\theta_1 = \theta_2$. It therefore suffices to restrict attention to elements of $\hT$ that satisfy $\theta_1 = \theta_2$. Further, note that the constraint is automatically satisfied if $\theta_1 = \theta_2 = 0$. So we may assume $\theta_2 > 0$.

Consider $\ut \in \hT$ such that $\theta_1 = \theta_2 \geq \theta_3 \geq \dots \geq \theta_n$, and $\theta_2 > 0$. Define $\hat{\ut} = \ut / \theta_2$; then $\hat{\ut} \in \hT$ with $1 = \hat{\theta}_1 = \hat{\theta}_2$. The left-hand side of (F) is homogeneous of order 1 in $\ut$. By our choice $\vas(a_i, \theta_i) = \theta_i U(a_i)$, the right-hand side of (F) is also homogeneous of order 1. As a consequence
\begin{align}
  \sum_{i=2}^{n-1}c_{i}(i\theta_{i+1}+(n-i)\theta_{i}) \le p_S(\ut)
  & \Leftrightarrow
  \theta_2 \sum_{i=2}^{n-1} c_{i} (i\hat{\theta}_{i+1} + (n-i)\hat{\theta}_{i}) \le p_S(\theta_2 \cdot \hat{\ut}) \nonumber \\
  & \Leftrightarrow
  \sum_{i=2}^{n-1} c_{i} (i\hat{\theta}_{i+1} + (n-i)\hat{\theta}_{i}) \le \frac{ p_S(\theta_2 \cdot \hat{\ut}) } {\theta_2} \nonumber  \\
  & \Leftrightarrow
  \sum_{i=2}^{n-1} c_{i} (i\hat{\theta}_{i+1} + (n-i)\hat{\theta}_{i}) \le p_S(\hat{\ut}). \label{eqn:simplifiedF}
\end{align}
(F) now simplifies, after removing the hats in $\hat{\ut}$, to
\[
	\text{(F)}\quad \sum_{i=2}^{n-1}c_{i}(i\theta_{i+1}+(n-i)\theta_{i})\le p_S(\ut), \ \, \forall\ut \in \hT, \, \theta_1=\theta_2=1.
\]

We now simplify the (W) constraint. First note that when $\theta_1 = 0$, (W) is trivially satisfied since it is always the case that $\sigma_S(\ut) \ge p_S(\ut)$. So we may assume that $\theta_1 > 0$. We next note that the summation $\sum_{i=2}^{n-1}c_{i}(i\theta_{i+1}+(n-i)\theta_{i})$, $\sigma_S(\ut)$, and $p_S(\ut)$ are all homogeneous of order 1 in $\ut$. Under $\theta_1 > 0$, we can re-scale $\ut$ by its first component to get $\hatt = \ut/\theta_1$, and obtain
\begin{eqnarray}
\lefteqn{\sum_{i=2}^{n-1}c_{i}(i\theta_{i+1}+(n-i)\theta_{i}) + t \sigma_S(\ut) \ge p_S(\ut)}\nonumber
\\ &\Leftrightarrow& ~~ \sum_{i=2}^{n-1}c_{i}(i\hat{\theta}_{i+1}+(n-i)\hat{\theta}_{i}) + t \sigma_S(\hatt) \ge p_S(\hatt). \label{eqn:simplifiedW}
\end{eqnarray}
Thus the worst-case constraint (W) simplifies to
\[
  \text{(W)}\quad  \sum_{i=2}^{n-1}c_{i}(i\theta_{i+1}+(n-i)\theta_{i}) + t \sigma_S(\ut) \ge p_S(\ut), \ \, \forall \ut \in \hT, \, \theta_1=1.
\]
We note that the simplification of constraints (F) and (W) was facilitated by our assumption of the surrogate valuation function $\vas(a,\theta)=\theta U(a)$. These simplifications, as we will observe in the proof of Theorem \ref{thm:verify-123}, enable us to obtain a simpler solution to the optimization problem at hand.

\subsection{An uncertain convex program}
\label{subsec:ucp}

Let us now define $\Theta = \{ \ut \in \hT : 1 = \theta_1 \}$. In view of the simplifications of (F) and (W), we can now rewrite the optimization problem as
\begin{alignat}{2}
        \label{eqn:final-opt-problem}
		& \min_{\uc,t} \ t \\
		\text{subject to } \hspace*{.3in}
		&\text{(F)}  \quad\, \sum_{i=2}^{n-1}c_{i}(i\theta_{i+1}+(n-i)\theta_{i})\le p_S(\ut), \ \forall\ut \in \Theta, \theta_2 = 1, \nonumber \\
		&\text{(VP)} \,\, \sum_{i=2}^k c_i \geq 0, \ k = 2, 3, \ldots, n-1, \nonumber \\
		&\text{(W)}  \,\,\,\, \sum_{i=2}^{n-1}c_{i}(i\theta_{i+1}+(n-i)\theta_{i}) + t \sigma_S(\ut) \ge p_S(\ut), \ \forall \ut \in \Theta. \nonumber
\end{alignat}

This optimization problem continues to be a generalized LP with a continuum of constraints. However, the continuum of constraints are now parametrized by a compact set $\Theta$ instead of the non compact set $\hT$.

\citet{Chorppath} studied the simpler VCG setting and adopted a randomized approach to solving the optimization problem within a probably approximately correct framework. Specifically, they considered a random sampling of constraints from $\Theta$ and provided guarantees on the number of samples required to obtain a near optimum solution. Here we take a deterministic approach.

\section{An uncertain convex program}
\label{sec:UCP}

The optimization in (\ref{eqn:final-opt-problem}) can be cast as a convex optimization problem subject to convex constraints having an uncertainty parameter. Such problems are called Uncertain Convex Programs (UCP). Formally, a UCP is defined (in \citet{ref:Calafiore03}) as a convex program of the form
\begin{equation}
  \label{eqn:ucp}
  \min_{x}f(x)\hspace*{.3in}\text{subject to}\hspace*{.3in}x \in A \text{ and } g(x,\ut)\leq 0,\forall\ut\in\Theta,
\end{equation}
where $\ut$ is the uncertainty parameter, $\Theta$ is an $n$-dimensional set, $x$ is a $d$-dimensional variable over which optimization occurs, $g(x,\ut)$ is a convex function of $x$ for every $\ut\in\Theta$, and $A$ is a convex subset of $\R^d$. In general, the index set for the constraints, $\Theta$, may be a continuum and, hence, the constraint set may be hard to characterize or compute.

The following three approaches to solve a UCP are known: robust optimization, chance-constrained optimization, and sampled convex program (SCP) \citep[see][]{ref:Calafiore03}. The first two techniques have been extensively studied under some special settings while the third technique (SCP) appears to have a wider applicability (see \citet{ref:Calafiore03,ref:Calafiore04} for details). The SCP technique involves sampling a subset $\hat{\Theta}^{(m)}=\{\ut^{(1)},\ut^{(2)},\ldots,\ut^{(m)}\}$ in an independent and identically distributed fashion according to a distribution $P_D$, and relaxing the constraining parameters from `$\ut\in\Theta$' to `$\ut\in\hat{\Theta}^{(m)}$'. Formally, an SCP is of the form
\begin{equation}
  \label{eqn:SCP}
  \min_{x}f(x)\hspace*{.3in}\text{subject to}\hspace*{.3in}x \in A \text{ and } \{g(x,\ut^{(i)})\leq 0,i=1,2,\ldots,m\},
\end{equation}
where $\ut^{(i)}$ is the $i^{\text{th}}$ sample of the constraint parameter, and $m$ is the number of samples. \citet{ref:Calafiore04} provide the number of samples $m$ sufficient to make the sampled constraint set approximate the actual constraint set in a particular sense as described next.

\begin{thm}
\citep{ref:Calafiore04}
\label{res:1}
Let the violation probability $V(x)$ at $x$ be defined as $V(x)=P_D(\ut\in\Theta:g(x,\ut)>0)$. Then, for a fixed $\epsilon,\delta>0$, the number of samples
$$
  m(\epsilon,\delta)=\frac{2}{\epsilon}\left(n\log{\left(\frac{2}{\epsilon}\right)}+\log{\left(\frac{1}{\delta}\right)}\right)+2n
$$
suffices for having $\Pr\{V(x)\leq\epsilon\}\geq 1-\delta$ for each $x$ that satisfies all the constraints of the SCP.
\end{thm}

In this paper, we want to bound the number of samples $m(\tau)$ needed to make the {\em values} of the UCP and the SCP be within $\tau$ of each other, that is,
\begin{equation}
  \label{eqn:value-tau-close}
  \mid\text{Value of SCP}-\text{Value of UCP}\mid\leq\tau.
\end{equation}
\citet{Chorppath}, following \citet{ref:Calafiore04}, studied a random sampling of constraints and provided a bound on the number of samples needed to satisfy (\ref{eqn:value-tau-close}) with high probability. Here we take a deterministic approach. We first state and prove a more general result to bring out the essential ideas. We then specialize it to the almost budget balance problem.

\subsection{Solution to a general UCP}
\label{subsec:solution-UCP}

An $\varepsilon$-cover for $\Theta$ is a collection of points $\hT^{(m)} = \{ \ut^{(1)}, \ut^{(2)}, \ldots, \ut^{(m)} \}$ such that balls of radius $\varepsilon$ centered around each of these points cover $\Theta$. If $\Theta$ is compact, there exists a finite cover.

Let $\X$ denote the constraint set of the UCP, and let $\Y$ denote the constraint set of the SCP obtained from an $\varepsilon$-cover. Symbolically,
\[
  \X = \bigcap_{\ut\in\Theta}\{x \in A :g(x,\ut)\leq 0\}\text{ and }\Y = \bigcap_{\ut\in\hat{\Theta}^{(m)}}\{x \in A :g(x,\ut)\leq 0\}.
\]
Since $\hat{\Theta}^{(m)} \subset \Theta$, we have $\X \subset \Y$. We make the following assumption.

\begin{assumption}~
\label{assumption-ucp}
\begin{itemize}
 \item[(a)] The mapping $x\mapsto f(x)$ is Lipschitz on $\Y$ with Lipschitz constant $K_1$.
 \item[(b)] The mapping $\ut \mapsto g(x,\ut)$ is uniformly Lipschitz over $x \in \Y$, with Lipschitz constant $K_2$.
 \item[(c)] There is a constant $K_3$ such that, for every $y \in \Y \setminus \X$, there exists a $\ut \in \Theta$ and an $x \in \X$ that satisfy $g(y,\ut) \geq K_3^{-1} ||y-x||$.
\end{itemize}
\end{assumption}

Our general result is the following.

\begin{thm}
\label{thm:0}
For the UCP (\ref{eqn:ucp}) and the SCP (\ref{eqn:SCP}) with $\hT^{(m)}$ being an $\varepsilon$-cover for $\Theta$, let Assumption \ref{assumption-ucp} hold. Then the optimal values of the UCP and the SCP are within $K_1 K_2 K_3 \varepsilon$ of each other.
\end{thm}

\begin{pf}  Let $x^*$ solve the UCP and let $y^*$ solve the SCP. Since $\X \subset \Y$, we have $f(x^*) \geq f(y^*)$. We may assume that $f(x^*) > f(y^*)$ (hence $y^* \in \Y \setminus \X$), for otherwise, the theorem is trivially true.

By Assumption \ref{assumption-ucp}(c), for this $y^*$, there exist a $\ut \in \Theta$, $x \in \X$, such that
\begin{equation}
	\label{eqn:K3}
	g(y^*,\ut) \geq K_3^{-1} \|y^*-x\|.
\end{equation}
Clearly, we must have $\ut \notin \hT^{(m)}$, for otherwise, $g(y^*, \ut) \leq 0$, which contradicts (\ref{eqn:K3}). Let $\ut^*$ be the element in the $\varepsilon$-cover $\hT^{(m)}$ that is closest to $\ut$. We then have $||\ut^* - \ut|| \leq \varepsilon$, and since $y^*$ is a feasible point for the SCP, we also have $g(y^*, \ut^*) \leq 0$. Thus
\begin{equation}
     \label{eqn:K2}
     g(y^*,\ut) \leq  g(y^*,\ut)-g(y^*,\ut^*)  \stackrel{(\star)}{\leq} K_2 \|\ut-\ut^*\| \leq K_2 \varepsilon,
\end{equation}
where ($\star$) follows from Assumption \ref{assumption-ucp}(b). Since $x \in \X$, we must also have $f(x^*) \leq f(x)$, and thus
$$
  f(y^*)\leq f(x^*)\leq f(x).
$$
Subtracting $f(y^*)$ throughout, we have,
\begin{equation}
  \label{eqn:K1}
  0 \leq f(x^*)-f(y^*) \leq f(x)-f(y^*) \leq K_1 \|y^*-x\|,
\end{equation}
where the last inequality follows from Assumption \ref{assumption-ucp}(a). Putting the chain of inequalities in (\ref{eqn:K3}), (\ref{eqn:K2}), and (\ref{eqn:K1}) together, we get $|f(x^*) - f(y^*)| \leq K_1 K_2 K_3 \varepsilon$.\qed
\end{pf}

\section{An application to the problem of almost budget balance}
\label{sec:apply-UCP}

The optimization problem in (\ref{eqn:final-opt-problem}) can be cast as a UCP. Let us see how.

Define $x_i = \sum_{j=2}^i c_j, \ i = 2, \ldots, n-1$, and $x_n = t$. (There is no $x_1$). Let $x = (x_2, \ldots, x_{n-1}, x_n)$.

(VP) now becomes $x_i \geq 0, i = 2, \ldots, n-1$. The variable $t$ is nonnegative, and hence $x_n \geq 0$. Moreover, (W) is trivially satisfied for $x_n \geq 1$, when $x_i \geq 0, i = 2, \ldots, n-1$. Therefore, restricting $x_n$ to be at most $1$ has no effect on the optimization problem since we minimize $x_n$. Thus the set $A$ for the UCP is
$$
  A=\{x ~|~ x_n \leq 1, x_i \geq 0, i = 2, \ldots, n\}.
$$

We now write (F) and (W) in terms of the above-defined variables. To do this, we define
\begin{eqnarray}
  \label{eqn:alpha-i}
  \alpha_i(\ut) & = & i \theta_{i+1} + (n-i) \theta_i, \,\, i = 2, \ldots, n-1 \\
  \label{eqn:alpha-n}
  \alpha_n(\ut) & = & 0.
\end{eqnarray}
Using $c_2 = x_2$ and $c_i = x_i - x_{i-1}$ for $i = 3, \ldots, n-1$, the (F) constraint then becomes
\[
  g_1(x, \ut) := \sum_{i=2}^{n-1} x_i (\alpha_i(\ut) - \alpha_{i+1}(\ut)) - p_S(\ut) \leq 0, \quad \forall \ut \in \Theta.
\]
Further, (W) becomes
\[
  g_2(x,\ut) :=  - \sum_{i=2}^{n-1} x_i (\alpha_i(\ut) - \alpha_{i+1}(\ut)) + p_S(\ut) - x_n \sigma_S(\ut) \leq 0, \quad \forall \ut \in \Theta.
\]
Now set $g(x,\ut) := \max \{ g_1(x,\ut), g_2(x,\ut) \}$. We earlier argued that we could restrict (F), $g_1(x,\ut) \leq 0$, to those $\ut \in \Theta$ that satisfy $\theta_2 = 1$. In order to combine the two constraints into a single one, we allow other values of $\theta_2$, $\theta_2\leq 1$, even though the constraints are tightest when $\theta_2 = 1$.

Finally, the objective function is taken to be $f(x) = x_n$. The UCP in (\ref{eqn:final-opt-problem}) is then of the form (\ref{eqn:ucp}) studied in the previous section.

We now have the following result.
\begin{thm}
\label{thm:verify-123}
With $\Theta$, $A$, $f$, $g$ as defined above, the corresponding UCP satisfies Assumption \ref{assumption-ucp}.
\end{thm}

\begin{pf}  See \ref{app:proof-of-123}.\qed
\end{pf}

By Theorems \ref{thm:0} and \ref{thm:verify-123}, using an $\varepsilon$-cover for $\Theta$, the values of the UCP for the almost budget balance problem in (\ref{eqn:final-opt-problem}) and the associated SCP are within $K_1 K_2 K_3 \varepsilon$ of each other. The proof of Theorem \ref{thm:verify-123} provides more information on how the constants $K_1$, $K_2$, and $K_3$ depend on $n$ and $U$. The proof, especially equations (\ref{eqn:g2-difference}) and (\ref{eqn:B2}), requires $\sigma_S(\ut)$ and $p_S(\ut)$ to be lower bounded by a positive number. The simplifications in (F) and (W) were needed to obtain these lower bounds.

\section{Discussion}
\label{sec:discussion}

\subsection{Summary}
We considered the allocation of a single divisible good among $ n $ agents whose valuation functions are private information known only to the respective agents. The social planner announces (1) an allocation scheme and a payment scheme that depend only on a {\em scalar bid} from each agent, and (2) invites the agents to submit their bids. Allocations and payments utilize a surrogate valuation function chosen (and announced beforehand) by the social planner (SSVCG). Rebates are used to achieve almost budget balance. We provided a framework to design the rebates and to achieve almost budget balance in a certain worst-case sense (\ref{eqn:ordered-opt-prob}). Our framework involved a solution to a convex optimization problem with a continuum of constraints for which we proposed a solution method involving constraint sampling. The almost budget balance property and the implementability of the mechanism holds off-equilibrium as well.

\subsection{Performance of linear rebates}
\label{subsec:surrogate}
Linear rebates are known to be optimal in the homogeneous discrete goods setting (\citet{GuoConitzer2010} and \citet{Moulin}). This was our main motivation for studying linear rebate functions in this paper. While the optimality of linear rebates in our divisible goods setting is not yet established in generality, we present some numerical results to highlight the reduction in budget surplus using linear rebates.

For our simulations, we chose the surrogate valuation function $\vas(a_i,\theta_i) = \theta_i U(a_i)$. This form of surrogate valuations is popular in the computer networking literature. (See \citet{Kelly97} for an example.) This choice is 1-homogeneous in the $\theta_i$ variable, a property which in conjunction with the choice of linear rebates and the scaling property of VCG payments enabled us to compactify the set of $\ut$'s to $\Theta$, the set of all ordered $\ut$'s with $\theta_1 = 1$.

We chose $U(a)=a^{1-\alpha}$, $\alpha\in\{0.01,0.25,0.5,0.75,0.99\}$. These $U$ functions are related to the generalized $\alpha$-logarithm suggested by \citet{YangHajek}. In each case, the coefficients of the linear rebate functions were obtained by solving the sampled convex problem (SCP). But instead of using an $\varepsilon$-cover, we used the $\ue_k$ profiles and $5000\times n$ additional randomly sampled constraints\footnote{$\theta_2,\ldots,\theta_n$ were picked uniformly at random from [0,1] and were then sorted.}. The corresponding value of the SCP is denoted ``Numerical'' value. An additional $50,000\times n$ samples were generated, and the performance of the identified rebate function on those $50,000\times n$ samples is denoted ``Simulated'' value. Since the rebates are determined only as an approximate solution with the sampled constraints, the worst-case ratio in the simulations, ``Simulated'' value, can be higher than ``Numerical'' value.

\begin{figure}
\begin{center}
\begin{tabular}{cc}
\subfloat[]{\label{sqrt_figure}\includegraphics[scale=0.3]{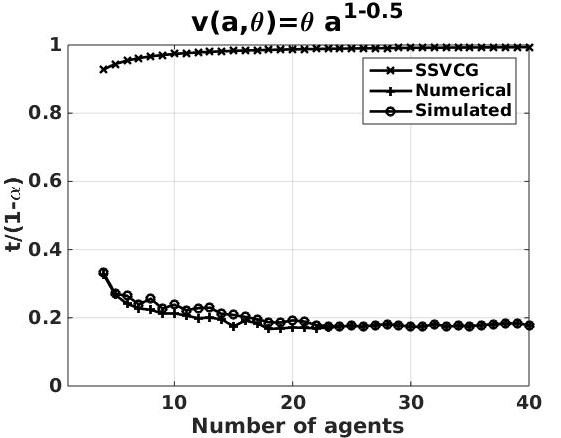}}&
\subfloat[]{\label{consolidate_figure}\includegraphics[scale=0.3]{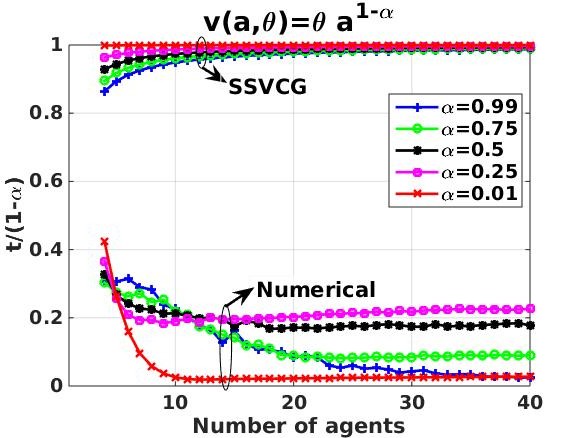}}
\end{tabular}
\caption{(a)(Worst-case objective $t$)/($1/2$) vs. number of agents; $\vas(a,\theta) = \theta\sqrt{a}$. (b) (Worst-case objective $t$)/($1-\alpha$) vs. number of agents; $\vas(a,\theta) = \theta a^{1-\alpha}$ for $\alpha\in\{0.01,0.25,0.5,0.75,0.99\}$. `Simulated' (not plotted) was close to `Numerical' in each case.}
\end{center}
\end{figure}

For $\alpha=0.5$, Figure \ref{sqrt_figure} shows ``Simulated'' and ``Numerical'' values and compares them with the ``SSVCG'' value, the surplus under no rebates. Figure \ref{consolidate_figure} shows ``Numerical'' and ``SSVCG'' values for each value of $\alpha$. ``Simulated'' (not plotted) and ``Numerical'' values were close to each other for each value of $\alpha$, and both significantly lower than ``SSVCG''. Moreover the worst-case ratio reduces as the number of agents increases. In contrast, the ``SSVCG'' value is nearly constant across the number of agents. Both of these plots provide a compelling argument in favor of linear rebates.

Observe that the worst-case objective in the plots is scaled by a factor of $1/(1-\alpha)$. This is because the observed SSVCG value was at most $1-\alpha$. This suggests (correctly) that we should set $\alpha$ very close to $1$ to reduce the worst-case objective. The reduction in the worst-case objective occurs because of an increase in $\sigma_S(\ut)$ at the Nash equilibrium point corresponding to each $\alpha$, and not because of the reduced surplus. The reduced objective function value is because of our choice of the Moulin objective function which reweighs the fraction of surplus retained by a factor $p_S(\ut) / \sigma_S(\ut)$. The surplus itself approaches a nonzero constant as the number of agents goes to infinity. (Proofs of these assertions can be found in the online appendix.) Interestingly, linear rebates continue to provide a significant reduction in the budget surplus, as can be gleaned from Figure \ref{consolidate_figure}. See Section \ref{subsec:extension} on possible extensions for further remarks.

The following is a brief description of how $|c_i|$, obtained from our simulations, vary in $i$, $n$, and $\alpha$.
\begin{itemize}
 \item $|c_2|$ is found to be the highest among all $|c_i|$. We did not observe any other increasing or decreasing trend in the variable $i$.
 \item $|c_i(n)|$ monotonically decreases in $n$ when $i\ll n$. $|c_i(n)|$ did not show any trend for $i$ comparable to $n$.
 \item $|c_2(\alpha)|$ was found to increase monotonically with $\alpha$. $|c_i(\alpha)|$ for all other $\alpha$ was found to increase up to some $\alpha$ and then decrease thereafter.
\end{itemize}
We reiterate that these are mere observations from simulation outcomes and are not formally established.

\subsection{Optimality in expectation}

In another work, \citet{GuoConitzer2010} considered a Bayesian setting and an associated objective: minimize the ratio $\mathbb{E}[$budget surplus$]/\mathbb{E}[$VCG payment without rebates$]$. Its extension to our setting involves a prior distribution on the space of valuation function profiles. Suppose one can assume that the Nash equilibrium associated with any valuation function profile is unique -- \citet{YangHajek} identify some sufficient conditions for this to hold. Then each valuation profile maps to a unique equilibrium bid profile. The prior distribution on the space of valuation function profiles induces a distribution on the space of equilibrium bid profiles. Now, taking expectations of the budget surplus with and without rebates, we can arrive at an objective function similar to that of \citet{GuoConitzer2010}.

\subsection{Possible extensions}
\label{subsec:extension}
As our choice of the almost budget balance criterion, we adopted the one proposed by \citet{Moulin}, see (\ref{eqn:Moulin-wc-abb-ssvcg}), for the reason highlighted in Section \ref{subsec:des-for-worst-case}. One could however work with the \citet{Guo} criterion, see (\ref{eqn:guo-conitzer-proposal}), and arrive at a new UCP. For this modified UCP, we do not have bounds on the size of the $\varepsilon$-cover because Assumption \ref{assumption-ucp}(c) does not hold.

We focused here on anonymous rebates because we assumed that the information with the social planner on the agents is symmetric to agent permutation. If information on the agents is asymmetric, for example the social planner wishes to weigh the allocation to one agent a little more than that to another, then agent-specific $U_i$ or agent-specific rebates could be used. Anonymity will have to be relaxed.

We restricted our attention to the allocation of an infinitely divisible good, and did not study a network setting, primarily because of our focus on identifying a suitable notion of almost budget balance and our desire to address the (VP) constraint's dependence on the true valuations in a simple setting. With our foundation, an extension to the network setting \citep{YangHajek} or a general convex setting \citep{Johari09} should now be possible.

While designing the rebate functions, we restricted attention to linear rebates for analytical tractability. Linear rebates are optimal in the homogeneous discrete goods setting. Proving the optimality or sub-optimality of linear rebates in our setting  could be a direction for future work.

\appendix
\section*{APPENDIX}

\section{Monotonicity and scaling properties of VCG payments}
\label{app:monotonicity}

We now prove monotonicity and scaling properties of the surplus under Clarke's payment rule, a result that may be of independent interest.

We consider a slightly more general setting than that of the paper. We now allow the surrogate valuation function to possibly depend on the agent and replace $\vas$ by $\vas_i$ for agent $i$. We also relax the restriction that $\vas_i(a_i,\theta_i)=\theta_i U(a_i)$, and make the following assumptions on the family $\vas_i$, $i = 1, 2, \ldots, n$.

\begin{assumption}~
\label{ass_monotonicity}
\begin{itemize}
  \item[(a)] For every $i$, every $\theta_i > 0$, $\vas_i(\cdot,\theta_i)$ is strictly concave, strictly increasing, continuously differentiable over $[0,1]$.
  \item[(b)] For every $a_i \in [0,1]$, the map $\theta_i \mapsto \vas_i(a_i,\theta_i)$ is absolutely continuous.
  \item[(c)] For every $a_i \in [0,1]$, the partial derivative $\frac{\partial \vas_i(a_i,\theta_i)}{\partial \theta_i}$ exists.
  \item[(d)] Furthermore, for some integrable $B_i(\theta_i)$, we have $\left| \frac{\partial \vas_i(a_i,\theta_i)}{\partial \theta_i} \right| \leq B_i(\theta_i).$
  \item[(e)] For each fixed $\theta_i$, the map $a_i \mapsto \frac{\partial \vas_i(a_i,\theta_i)}{\partial \theta_i}$ is increasing.
  \item[(f)] For each fixed $a_i$, the map $\theta_i \mapsto \frac{1}{\theta_i} \frac{\partial \vas_i(a_i,\theta_i)}{\partial a_i}$ is decreasing.
\end{itemize}
\end{assumption}

Obviously, $\vas_i(a_i, \theta_i) = \theta_i U(a_i)$, where $U$ is strictly concave, strictly increasing, and continuously differentiable over $[0,1]$ satisfies Assumption \ref{ass_monotonicity}.

The surplus under the Clarke's payment rule is given by
\[
p_S(\ut) = -(n-1) \sum_{j=1}^n \vas_j(a_j^*,b_j) + \sum_{i=1}^{n} \sum_{j \ne i} \vas_j(a_{-i,j}^*,b_j).
\]
See also (\ref{eqn:p-S}). The optimal social welfare is (see (\ref{eqn:sigma-S})):
\[
\sigma_S(\ut) = \sum_{i=1}^n \vas_i(a_i^*,\theta_i).
\]
The following shows the intuitive property that $p_S(\theta_i, \ut_{-i})$ is increasing in $\theta_i$.

\begin{prop}
\label{prop:monotone-scaling}
Under Assumption \ref{ass_monotonicity}, $p_S(\ut)$ satisfies following properties.\\
(a) ({\bf Monotonicity}) For fixed $\ut_{-i}$, the map $\theta_i \mapsto p_S(\theta_i, \ut_{-i})$ is increasing.\\
(b) ({\bf Scaling}) For fixed $\ut$, the map $\lambda \mapsto p_S(\lambda \ut) / \lambda$ is decreasing.
\end{prop}

\begin{pf}  (a) Fix $i$. The proof uses the envelope theorem \citep[see][Thm. 2]{milgrom-segal-2002}. Focus first on $\sigma_S$. By virtue of Assumption \ref{ass_monotonicity}(b)-(d), for a fixed $\ua$ and $\ut_{-i}$, we have that $\theta_i\mapsto\sum_{k=1}^n \vas_k(a_k, \theta_k)$ is absolutely continuous, has partial derivative, and
\[
  \left| \frac{\partial}{\partial \theta_i} \left( \sum_{k=1}^n \vas_k(a_k,\theta_k) \right) \right| = \left| \frac{\partial \vas_i(a_i,\theta_i)}{\partial \theta_i} \right| \leq B_i(\theta_i).
\]
By \citep[Thm. 2]{milgrom-segal-2002}, $\sigma_S(\ut)$ has a partial derivative with respect to $\theta_i$ almost everywhere on $[0,1]$ which equals
\[
  \frac{\partial \sigma_S(\ut)}{\partial \theta_i} =  \frac{\partial \vas_i(a_i^*(\ut),\theta_i)}{\partial \theta_i}.
\]
For each $k \neq i$, apply the same argument as given above to the envelope $\sum_{j \neq k} \vas_j(a^*_{-k,j}(\ut_{-k}), \theta_j)$, to get
\[
  \frac{\partial}{\partial \theta_i} \left( \sum_{j \neq k} \vas_j(a^*_{-k,j}(\ut_{-k}), \theta_j) \right) =  \frac{\partial \vas_i(a_{-k,i}^*(\ut_{-k}),\theta_i)}{\partial \theta_i}.
\]
For $k = i$, the corresponding envelope does not depend on $\theta_i$. The above considerations yield
\begin{equation}
  \frac{\partial p_S(\ut)}{\partial \theta_i}  =  \sum_{k \ne i} \frac{\partial \vas_i(a_{-k,i}^*(\ut_{-k}),\theta_i)}{\partial \theta_i}
 - (n-1) \frac{\partial \vas_i(a_i^*(\ut),\theta_i)}{\partial \theta_i}.
  \label{eqn:partial-p-V}
\end{equation}
It is easy to see (using the Karush-Kuhn-Tucker necessary conditions) that for each $k$, we have $a^*_{-k,i}(\ut_{-k}) \geq a^*_i(\ut)$. Intuitively, if an agent $k \neq i$ is out of consideration, then the optimal amount allocated to agent $i$ only increases. Consequently, by Assumption \ref{ass_monotonicity}(e), we have
\[
  \frac{\partial \vas_i(a_{-k,i}^*(\ut_{-k}),\theta_i)}{\partial \theta_i} \geq \frac{\partial \vas_i(a_i^*(\ut),\theta_i)}{\partial \theta_i}.
\]
Using this in (\ref{eqn:partial-p-V}), we get $\frac{\partial p_S(\ut)}{\partial \theta_i} \geq 0$, i.e., $p_S(\ut)$ is increasing in $\theta_i$.

(b) Differentiating $p_S(\lambda \ut) / \lambda$ with respect to $\lambda$, we get
\[
  \frac{d}{d\lambda}\left(\frac{p_S(\lambda\ut)}{\lambda}\right)=\frac{\lambda \ut^T \nabla_{\ut} p_S(\lambda\ut)-p_S(\lambda\ut)}{\lambda^2}.
\]
It suffices to show that this is negative. Without loss of generality, we can replace $\lambda \ut$ by $\ut$, and it suffices to check that $\ut^T \nabla_{\ut} p_S(\ut) \leq p_S(\ut)$. Using the formula (\ref{eqn:partial-p-V}), this amounts to checking that
\begin{eqnarray*}
 \lefteqn{ \sum_{i=1}^n \theta_i \left\lbrace\sum_{k \ne i} \frac{\partial}{\partial\theta_i}\vas_i(a^*_{-k,i}(\ut_{-k}),\theta_i)-(n-1)\frac{\partial}{\partial\theta_i}\vas_i(a^*_i(\ut),\theta_i)\right\rbrace} \\
&  \leq & \sum_{i=1}^n\left\lbrace\sum_{k\ne i}\vas_i(a^*_{-k,i}(\ut_{-k}),\theta_i)-(n-1)\vas_i(a^*_i(\ut),\theta_i)\right\rbrace.
\end{eqnarray*}
This holds if, for every $i$ and every $k \neq i$,
\begin{equation*}
  \theta_i\frac{\partial}{\partial\theta_i}\vas_i(a^*_{-k,i}(\ut_{-k}),\theta_i)-\theta_i\frac{\partial}{\partial\theta_i}\vas_i(a^*_i(\ut),\theta_i)
  \leq \vas_i(a^*_{-k,i}(\ut_{-k}),\theta_i)-\vas_i(a^*_i(\ut),\theta_i),
\end{equation*}
or equivalently
\begin{equation*}
  \theta_i\frac{\partial}{\partial\theta_i}\vas_i(a^*_{-k,i}(\ut_{-k}),\theta_i) - \vas_i(a^*_{-k,i}(\ut_{-k}),\theta_i)
  \leq \theta_i\frac{\partial}{\partial\theta_i}\vas_i(a^*_i(\ut),\theta_i) - \vas_i(a^*_i(\ut),\theta_i).
\end{equation*}
But this follows from $a^*_{-k,i}(\ut_{-k}) \geq a^*_i(\ut)$ and the fact that, for every $\theta_i$,
\[
  a_i \mapsto \theta_i\frac{\partial}{\partial\theta_i}\vas_i(a_i,\theta_i) - \vas_i(a_i,\theta_i)
\]
is decreasing, which is an easy consequence of Assumption \ref{ass_monotonicity}(f).\qed
\end{pf}

\section{Proof of Theorem \ref{thm:verify-123}}
\label{app:proof-of-123}

We again restrict the surrogate valuation function $\vas(a,\theta)=\theta U(a)$, with $U$ satisfying Assumption \ref{ass_surrogate}. We now verify (a)-(c) of Assumption \ref{assumption-ucp}.

Assumption \ref{assumption-ucp}(a): Since $f(x) = x_n$, this holds trivially with Lipschitz constant $K_1 = 1$.

Before we get to verifying the next assumption, we establish two lemmas.

\begin{lem}
\label{lem:positivity-x-coeff}
The coefficients of $x_i$ in the expression for $g_1$ are nonnegative, i.e., $\alpha_i(\ut) - \alpha_{i+1}(\ut) \geq 0$ for $i = 2, \ldots, n-1$, where $\alpha_i$ are defined in (\ref{eqn:alpha-i}) and (\ref{eqn:alpha-n}).
\end{lem}

\begin{pf}  For $i=2, \ldots, n-2$, using (\ref{eqn:alpha-i}), the coefficients of $x_i$ satisfy
  \[
    \alpha_i(\ut) - \alpha_{i+1}(\ut) = i(\theta_{i+1} - \theta_{i+2}) + (\theta_i - \theta_{i+2}) + (n-i-1) (\theta_i - \theta_{i+1}) \geq 0
  \]
where the last inequality follows because the $\theta_i$ are nonincreasing with index $i$. Finally, the coefficient of $x_{n-1}$ is simply $\alpha_{n-1}(\ut)$ which is nonnegative.\qed
\end{pf}

We next argue that the elements of $\X$ are bounded.

\begin{lem}
\label{lem:xi-bound}
  If $x \in \X$, then, for $i=2, \ldots, n-1$, we have $0 \leq x_i \leq B_n$, where $B_n := p_S(\ue_n)$.
\end{lem}

\begin{pf}  From (F) in (\ref{eqn:final-opt-problem}), the nonnegativity of $x_i$, and the nonnegativity of the coefficients of $x_i$ in the expression for $g_1$ established in Lemma \ref{lem:positivity-x-coeff}, we have that for each $i = 2, \ldots, n-1$,
\[
  x_i \leq \frac{p_S(\ut)}{\alpha_i(\ut) - \alpha_{i+1}(\ut)}, \forall \ut \in \Theta.
\]
Setting $\ut = \ue_i$, we get $\alpha_i(\ut) - \alpha_{i+1}(\ut) = (n-i) - 0$. Now, by using monotonicity of $p_S$ (Proposition \ref{prop:monotone-scaling}(a)), we get
\[
  x_i \leq \frac{p_S(\ue_i)}{n-i} \leq \frac{p_S(\ue_n)}{n-i} \leq p_S(\ue_n) = B_n.
\]
Hence the result.\qed
\end{pf}

We now continue with the proof that Assumption \ref{assumption-ucp}(b) holds.

Assumption \ref{assumption-ucp}(b): If $g_1(x, \cdot)$ and $g_2(x,\cdot)$ are both uniformly Lipschitz with constants $K_2'$ and $K_2''$, then $g(x,\cdot) = \max \{ g_1 (x,\cdot), g_2(x,\cdot) \}$ is also uniformly Lipschitz with constant $K_2 = \max\{ K_2', K_2''\}$. The mappings $g_1(x,\cdot)$ and $g_2(x,\cdot)$ are uniformly Lipschitz because of the following:
\begin{itemize}
  \item[(i)] $\ut \mapsto \sigma_S(\ut)$ is Lipschitz with constant $U(1)\sqrt{n}$;
  \item[(ii)] $\ut \mapsto p_S(\ut)$ is Lipschitz with constant $2U(1)n\sqrt{n}$;
  \item[(iii)] the mapping $\ut \mapsto \sum_{i=2}^{n-1}x_{i}(\alpha_i(\ut) - \alpha_{i+1}(\ut))$ is uniformly Lipschitz.
\end{itemize}
Items (i) and (ii) were established in Lemma \ref{lem:sigmasps-lipschitz}. To see (iii), observe that
\begin{multline*}
\left|\alpha_i(\ut) - \alpha_i(\ut')\right| =  \left|i(\theta_{i+1} - \theta_{i+1}') + (n-i)(\theta_{i} - \theta_{i}')  \right| \\
\leq i ||\ut - \ut'|| + (n-i) ||\ut - \ut'|| = n ||\ut - \ut'||.
\end{multline*}
Since this inequality holds for all $i$, and since $x_i \leq B_n$, we see that Item (iii) holds with Lipschitz constant $2 n^2 B_n$. The Lipschitz constants $K_2'$ and $K_2''$, and hence $K_2$, are as follows.
\begin{eqnarray*}
K_2' & = & 2n^2 B_n + 2U(1)n\sqrt{n}\\
K_2'' & = & 2n^2 B_n + 2U(1)n\sqrt{n} + U(1)\sqrt{n} \;\;\;\;\;\;\;\;\;\;\;(\mbox{from }x_n \leq 1)\\
K_2 & = & \max\{K_2', K_2''\} = K_2''.
\end{eqnarray*}

Assumption \ref{assumption-ucp}(c): Define $\X_i = \{ x \in A \, |\,  g_i(x,\ut) \leq 0 ~\forall \ut \in \Theta \}, i = 1,2$. Note that $g_1(x,\ut)$ does not depend on the last component (the $x_n$ component). Recall that $\Y$ is the constraint set for the SCP.

Let $y = (y_2, \cdots, y_{n-1}, y_n) \in \Y \setminus \X$. We consider two cases.

(i) Suppose $y \in \X_1$, but $y \notin \X_2$.

Find the smallest $t > y_n$ such that $x = (y_2, \ldots, y_{n-1}, t) \in \X_2$. Since the first $n-2$ components have not changed, $x \in \X_1$ as well, and so $x \in \X$. We now claim that there is a $\ut$ with $g_2(x,\ut) = 0$.

Suppose that the claim is false. Since $\ut\mapsto g_2(x,\ut)$ is continuous, and since $\Theta$ is compact, we have $-\varepsilon:=\max_{\ut\in\Theta}g_2(x,\ut)<0$. By the linearity of $g_2(.,\ut)$, for any $\ut\in\Theta$, we have
\begin{align*}
  g_2\left(\left(y_2,\ldots,y_{n-1},t-\frac{\varepsilon}{nU(1)}\right),\ut\right) & = g_2((y_2,\ldots,y_{n-1},t),\ut) + \frac{\varepsilon}{nU(1)}\sigma_S(\ut) \\ & \leq g_2(x,\ut) + \frac{\varepsilon}{nU(1)}\sigma_S(\ut) \\ & \leq -\varepsilon + \frac{\varepsilon}{nU(1)}\sigma_S(\ut) \leq 0,
\end{align*}
where $\sigma_S(\ut)\leq nU(1)$ since $\theta_i,a_i\in[0,1]\,\forall i$. Thus $(y_2,\ldots,y_{n-1},t-\frac{\varepsilon}{nU(1)})\in\X_2$, and this contradicts the choice of $t$.

Using the claim, we have
\begin{multline}
\label{eqn:ass-2c}
  g(y,\ut) \geq g_2(y,\ut) = g_2(y,\ut) - g_2(x, \ut) = (t-y_n) \sigma_S(\ut) = ||y - x|| \sigma_S(\ut) \\ \geq ||y - x|| U(1).
\end{multline}
The last inequality holds because $\sigma_S(\ut)$ is the socially optimum value when the reported bids are $\ut$, and $U(1)$ is the value obtained for a particular allocation that gives the entire good to a single agent. In (\ref{eqn:ass-2c}), we also used $||y-x|| = t-y_n$ because $y$ and $x$ differ only in the last component and $t > y_n$. Let $K_3' = U(1)$. We will choose $K_3^{-1}$ lower than this after considering other cases.

(ii) Suppose $y \notin \X_1$.

Fix $\gamma > n/U(1)$. We will choose this $\gamma$ suitably later. Find the smallest $\beta > 0$ such that
\[
  x(\beta) := [(y_2 - \beta)_+, \ldots, (y_{n-1} - \beta)_+, \min(1,y_n+\gamma\beta)] \in \X,
\]
where $[\cdot]_+$ indicates truncation from below by 0. Notice that the first $n-2$ components decrease, but the last component increases. The procedure clearly terminates because $(0, \ldots, 0, 1) \in \X$, and this point will eventually be reached for some $\beta > 0$.

We now have two subcases.

(ii-a): The path from $y$ to $x(\beta)$ first touches $\X_2$, say at $x(\beta')$ for $0 \leq \beta' \leq \beta$, before entering $\X$.

Then $x(\beta'') \in \X_2$ for all $\{\beta'' \geq \beta' ~|~ y_n + \gamma\beta'' \leq 1\}$. This is because for all $\ut \in \Theta$, we have
\begin{align}
  & g_2(x(\beta'), \ut) - g_2(x(\beta''), \ut) \nonumber \\
  & = -\sum_{i=2}^{n-1} (x_i(\beta') - x_i(\beta'')) (\alpha_i(\ut) - \alpha_{i+1}(\ut)) - \gamma(\beta' - \beta'') \sigma_S(\ut) \nonumber \\
  & \geq -(\beta'' - \beta') \sum_{i=2}^{n-1} (\alpha_i(\ut) - \alpha_{i+1}(\ut)) + \gamma(\beta'' - \beta') \sigma_S(\ut) \label{eqn:g2-difference-pre},
\end{align}
where (\ref{eqn:g2-difference-pre}) follows from $\alpha_i(\ut) - \alpha_{i+1}(\ut) \ge 0$ (Lemma \ref{lem:positivity-x-coeff}) and $x_i(\beta') - x_i(\beta'') \le \beta'' - \beta'$ for $i = 2, 3, \ldots, n-1$. Continuing, the right-hand side of (\ref{eqn:g2-difference-pre}) equals
\begin{align}
  (\beta'' - \beta') \left[ - (\alpha_2(\ut) - \alpha_n(\ut)) + \gamma \sigma_S(\ut) \right]
  & = (\beta'' - \beta') \left[ \gamma \sigma_S(\ut) - \alpha_2(\ut) \right] \nonumber \\
  \label{eqn:g2-difference}
  & \geq (\beta'' - \beta') \left[ \gamma U(1) - n \right] \\
  \label{eqn:g2-difference-lb}
  & \geq 0.
\end{align}
In (\ref{eqn:g2-difference}), we used $\sigma_S(\ut) \geq U(1)$ and $\alpha_2(\ut) \leq n$. The latter follows because $\alpha_2(\ut) = 2 \theta_3 + (n-2) \theta_2 \leq 2 + n-2 =n$, since $0 \leq \theta_i \leq 1$ for all $i$. The inequality (\ref{eqn:g2-difference-lb}) follows from the choice of $\gamma$.

In case $\beta'' \in [\beta',\beta]$ but $y_n + \gamma\beta'' > 1$, then $x_n(\beta'')=1$ and thus $x(\beta'') \in \X_2$ is trivially true.

We claim that there exists a $\ut \in \Theta$ with $\theta_2 = 1$ such that $g_1(x(\beta),\ut) = 0$. Suppose that the claim is false. Since $\ut\mapsto g_1(x,\ut)$ is continuous, and since $\Theta$ is compact, we have $-\varepsilon:=\max_{\{\ut\in\Theta:\theta_2=1\}}g_1(x,\ut)<0$. By the linearity of $g_1(.,\ut)$, for any $\ut\in\Theta$ with $\theta_2=1$, we have
\begin{multline}
  g_1\left(x\left(\beta-\frac{\varepsilon}{n}\right),\ut\right) \leq g_1(x(\beta),\ut) + \frac{\varepsilon}{n}\sum_{i=2}^{n-1}(\alpha_i(\ut)-\alpha_{i+1}(\ut)) \\ \leq -\varepsilon + \frac{\varepsilon}{n}(\alpha_2(\ut)-\alpha_n(\ut)) \leq 0,
\end{multline}
i.e., $x\left(\beta-\frac{\varepsilon}{n}\right)\in\X_1$, and this contradicts the choice of $\beta$.

Now for this $\ut$, which satisfies $\theta_1 = \theta_2 = 1$, by monotonicity of $p_S$, we have $p_S(\ut) \geq p_S(\ue_2) = 2U(1) - 2U(1/2):= B_2 > 0$. The strict positivity follows because $U$ is strictly increasing. We then have
\begin{multline}\label{eqn:g(y-ut)}
  g(y, \ut) \geq g_1(y,\ut) = g_1(y,\ut) - g_1(x(\beta),\ut)\\=\sum_{i=2}^{n-1} (y_i - x_i(\beta))(\alpha_i(\ut) - \alpha_{i+1}(\ut)) 
  \geq \beta \sum_{i=2}^{n-1} (\alpha_i(\ut) - \alpha_{i+1}(\ut)) {\bf 1}\{x_i(\beta) > 0\}.
\end{multline}
The last inequality follows because: if $x_i(\beta) > 0$, the difference $y_i - x_i(\beta)$ is exactly $\beta$; if not, the corresponding term is $\geq 0$ and thus can be dropped.

To lower bound this last term, we proceed as follows. Since $p_S(\ut) \geq p_S(\ue_2) = B_2$, and since $g_1(x(\beta),\ut) = 0$, we have
\begin{eqnarray}
  B_2 & \leq & p_S(\ut) = \sum_{i=2}^{n-1} x_i(\beta) (\alpha_i(\ut) - \alpha_{i+1}(\ut)) \;\;\;\;\;\;\;\;\;\;\;\;
  (\mbox{from~}g_1(x(\beta),\ut) = 0) \nonumber\\
      & = & \sum_{i=2}^{n-1} x_i(\beta) (\alpha_i(\ut) - \alpha_{i+1}(\ut)) {\bf 1}\{x_i(\beta) > 0\} \nonumber \\
      \label{eqn:B2}
      & \leq & B_n \sum_{i=2}^{n-1} (\alpha_i(\ut) - \alpha_{i+1}(\ut)) {\bf 1}\{x_i(\beta) > 0\},
\end{eqnarray}
where the last inequality follows from Lemma \ref{lem:xi-bound}. Putting (\ref{eqn:g(y-ut)}) and (\ref{eqn:B2}) together, we get
\begin{equation}
  \label{eqn:g(y-ut)-1}
  g(y, \ut) \geq \beta B_2 / B_n.
\end{equation}

Define $x^*(\beta) = y - \beta(1,\ldots,1,-\gamma)$. This is without the truncation from below by 0. Clearly,
\begin{equation}
  \label{eqn:norm-y-x}
  ||y - x(\beta)|| \leq ||y - x^*(\beta)|| = \beta ||(1,\ldots,1,-\gamma)|| = \beta \sqrt{n-2+\gamma^2}.
\end{equation}
Combining (\ref{eqn:g(y-ut)-1}) and (\ref{eqn:norm-y-x}), we get
\begin{equation}
  \label{eqn:g(y-ut)-2}
  g(y, \ut) \geq  \frac{B_2}{B_n \sqrt{n-2+\gamma^2}} ||y - x(\beta)||.
\end{equation}
Take $x = x(\beta)$ and take $K_3'' = B_2/\left(B_n \sqrt{n-2+\gamma^2}\right)$.

(ii-b): The path from $y$ to $x(\beta)$ first touches $\X_1$, say at $x(\beta')$ for $0 < \beta' < \beta$, before entering $\X$.

Then $x(\beta'') \in \X_1$ for all $\beta' \leq \beta'' \leq \beta$. This is because, once $\X_1$ is reached, the first $n-2$ components only decrease thereafter with $\beta$, and therefore $g_1$ also only decreases.

We now claim that there exists a $\ut \in \Theta$ such that $g_2(x(\beta),\ut) = 0$.  Suppose that the claim is false. Since $\ut\mapsto g_2(x,\ut)$ is continuous, and since $\Theta$ is compact, we have $-\varepsilon:=\max_{\ut\in\Theta}g_2(x,\ut)<0$. By the linearity of $g_2(.,\ut)$, for any $\ut\in\Theta$, we have
\begin{multline}
  g_2\left(x\left(\beta-\frac{\varepsilon}{\gamma nU(1)}\right),\ut\right) \leq g_2(x(\beta),\ut) + \frac{\varepsilon}{\gamma nU(1)} \gamma \sigma_S(\ut) \\ \leq -\varepsilon + \frac{\varepsilon}{nU(1)} \sigma_S(\ut) \leq 0,
\end{multline}
where first inequality follows since the terms involving $(x_2(\beta),\ldots,x_{n-1}(\beta))$ are nonpositive. Thus $x\left(\beta-\frac{\varepsilon}{\gamma nU(1)}\right)\in\X_2$, and this contradicts the choice of $\beta$. We then have
\begin{eqnarray}
  g(y,\ut) & \geq & g_2(y,\ut) = g_2(y,\ut) - g_2(x(\beta),\ut) \nonumber \\
  \label{eqn:caseii-b}
  & \geq & \beta(\gamma U(1) - n) \\
  \label{eqn:caseii-b-1}
  & \geq & \frac{(\gamma U(1) - n)}{\sqrt{n-2 + \gamma^2}} ||y - x(\beta)||
\end{eqnarray}
where (\ref{eqn:caseii-b}) follows by tracing the same sequence of inequalities leading to (\ref{eqn:g2-difference}), with $y$ and $x(\beta)$ in place of $x(\beta')$ and $x(\beta'')$, respectively, and (\ref{eqn:caseii-b-1}) follows from (\ref{eqn:norm-y-x}). Observe that the same sequence of inequalities leading to (\ref{eqn:g2-difference}) can be traced since $x(\beta) \in \X$ the moment $y_n + \gamma\beta = 1$, and thus $y_n + \gamma\beta > 1$ never occurs. Take $x = x(\beta)$ and $K_3''' = (\gamma U(1) - n)/\sqrt{n-2 + \gamma^2}$.

Setting $K_3^{-1} = \min\{ K_3', K_3'', K_3''' \}$, we see that Assumption \ref{assumption-ucp}-(c) holds. Setting $\gamma = (n + B_2/B_n)/U(1)$, which exceeds $n/U(1)$, we get $K_3^{-1} = \min\left\{ U(1), B_2/\left(B_n\sqrt{n - 2 + \gamma^2}\right)\right\}.$\qed

\section*{Acknowledgments}

This work was supported by the Department of Science and Technology [grant number SR/S3/EECE/0056/2011]; and by the Defence Research and Development Organisation [grant number DRDO0667] under the DRDO-IISc Frontiers Research Programme.

\section*{References}
\bibliographystyle{elsarticle-harv.bst}
\bibliography{refer1.bib}

\begin{thebibliography}{25}
\expandafter\ifx\csname natexlab\endcsname\relax\def\natexlab#1{#1}\fi
\expandafter\ifx\csname url\endcsname\relax
  \def\url#1{\texttt{#1}}\fi
\expandafter\ifx\csname urlprefix\endcsname\relax\def\urlprefix{URL }\fi

\bibitem[{Blumrosen et~al.(2007)Blumrosen, Nisan, and
  Segal}]{blumrosen2007auctions}
Blumrosen, L., Nisan, N., Segal, I., 2007. Auctions with severely bounded
  communication. J. Artif. Intell. Res.(JAIR) 28, 233--266.

\bibitem[{Calafiore and Campi(2005)}]{ref:Calafiore03}
Calafiore, G.~C., Campi, M.~C., 2005. Uncertain convex programs: {R}andomized
  solutions and confidence levels. Math. Prog. 102~(1), 25--46.

\bibitem[{Calafiore and Campi(2006)}]{ref:Calafiore04}
Calafiore, G.~C., Campi, M.~C., 2006. The scenario approach to robust control
  design. IEEE Trans. on Automatic Control 51~(5), 742--753.

\bibitem[{Cavallo(2006)}]{Cavallo06}
Cavallo, R., 2006. Optimal decision-making with minimal waste: strategyproof
  redistribution of {VCG} payments. In: Proceedings of the fifth international
  joint conference on Autonomous agents and multiagent systems. AAMAS '06. pp.
  882--889.

\bibitem[{Chorppath et~al.(2011)Chorppath, Bhashyam, and
  Sundaresan}]{Chorppath}
Chorppath, A.~K., Bhashyam, S., Sundaresan, R., July 2011. A convex
  optimization framework for almost budget balanced allocation of a divisible
  good. IEEE Trans. Autom. Sci. Eng. 8~(3), 520--531.

\bibitem[{Clarke(1971)}]{Clarke}
Clarke, E., 1971. Multipart pricing of public goods. Public Choice 2, 19--33.

\bibitem[{Green and Laffont(1977)}]{GreenLaffont1977}
Green, J., Laffont, J.-J., 1977. Characterization of satisfactory mechanisms
  for the revelation of preferences for public goods. Econometrica 45,
  427--438.

\bibitem[{Groves(1973)}]{Groves}
Groves, T., 1973. Incentives in teams. Econometrica 41~(4), 617--631.

\bibitem[{Gujar and Narahari(2009)}]{GujNar09}
Gujar, S., Narahari, Y., 2009. Redistribution mechanisms for assignment of
  heterogeneous objects. In: Formal Approaches Multi-Agent Systems, (FAMAS09).
  pp. 438--445.

\bibitem[{Gujar and Narahari(2011)}]{GujNar11}
Gujar, S., Narahari, Y., 2011. Redistribution mechanisms for assignment of
  heterogeneous objects. J. Artif. Intel. Res. (JAIR) 41, 131--154.

\bibitem[{Guo and Conitzer(2009)}]{Guo}
Guo, M., Conitzer, V., September 2009. Worst-case optimal redistribution of
  {VCG} payments in multi-unit auctions. Games and Economic Behavior 67~(1),
  69--98.

\bibitem[{Guo and Conitzer(2010)}]{GuoConitzer2010}
Guo, M., Conitzer, V., 2010. Optimal-in-expectation redistribution mechanisms.
  Artificial Intelligence 174~(5-6), 363 -- 381.

\bibitem[{Holmstr{\"o}m(1979)}]{Holmstrom79}
Holmstr{\"o}m, B., 1979. Groves' scheme on restricted domains. Econometrica
  47~(5), 1137--1144.

\bibitem[{Jain and Walrand(2010)}]{jain2010efficient}
Jain, R., Walrand, J., 2010. An efficient nash-implementation mechanism for
  network resource allocation. Automatica 46~(8), 1276--1283.

\bibitem[{Johari and Tsitsiklis(2004)}]{Johari2004}
Johari, R., Tsitsiklis, J.~N., 2004. Efficiency loss in a network resource
  allocation game. Mathematics of Operations Research 29~(3), 407--435.

\bibitem[{Johari and Tsitsiklis(2009)}]{Johari09}
Johari, R., Tsitsiklis, J.~N., 2009. Efficiency of scalar-parameterized
  mechanisms. Operations Research 57~(4), 823--839.

\bibitem[{Kakhbod and Teneketzis(2012)}]{kakhbod-2012}
Kakhbod, A., Teneketzis, D., 2012. An efficient game form for unicast service
  provisioning. IEEE Trans. on Automatic Ctrl. 57~(2), 392--404.

\bibitem[{Kelly(1997)}]{Kelly97}
Kelly, F., 1997. Charging and rate control for elastic traffic. European
  Transactions on Telecommunications 8~(1), 33--37.

\bibitem[{Milgrom and Segal(2002)}]{milgrom-segal-2002}
Milgrom, P., Segal, I., 2002. Envelope theorems for arbitrary choice sets.
  Econometrica 70~(2), 583--601.

\bibitem[{Moulin(2009)}]{Moulin}
Moulin, H., January 2009. Almost budget-balanced {VCG} mechanisms to assign
  multiple objects. Journal of Economic Theory 144~(1), 96--119.

\bibitem[{Reichelstein and Reiter(1988)}]{reichelstein1988game}
Reichelstein, S., Reiter, S., 1988. Game forms with minimal message spaces.
  Econometrica: Journal of the Econometric Society, 661--692.

\bibitem[{Semret(1999)}]{semret1999market}
Semret, N., 1999. Market mechanisms for network resource sharing. Ph.D. thesis,
  Columbia University.

\bibitem[{Sinha and Anastasopoulos(2013)}]{sinha-2013}
Sinha, A., Anastasopoulos, A., 2013. Generalized proportional allocation
  mechanism design for multi-rate multicast service on the internet. In:
  Communication, Control, and Computing (Allerton), 2013 51st Annual Allerton
  Conference on. IEEE, pp. 146--153.

\bibitem[{Vickrey(1961)}]{Vickrey}
Vickrey, W., 1961. Counterspeculation, auctions, and competitive sealed
  tenders. The Journal of Finance 16~(1), 8--37.

\bibitem[{Yang and Hajek(2007)}]{YangHajek}
Yang, S., Hajek, B., 2007. {VCG}-{K}elly mechanisms for allocation of divisible
  goods: Adapting {VCG} mechanisms to one-dimensional signals. IEEE Journal on
  Selected Areas in Communications 25~(6), 1237--1243.

\end{thebibliography}

\newpage
\section*{Online Appendix}

\textbf{Proof of Lemma \ref{NE is Full Rn}:}

Fix $\ut \in (0,\infty)^n$. Suppose $v_i(a_i) = \vas(a_i, \theta_i)$ $\forall i$. The valuation function $v_i$ inherits the concavity, strictly increasing, and continuously differentiable properties from $\vas$. All agents have infinite marginal utility at zero. Assumption \ref{ass_efficient_NE}(a) is thus clearly satisfied. Agent $i$ does not know the valuation functions of the other agents, but she realizes that there is a bid value that will make her resulting utility (after payment) align with the function optimized by the social planner, regardless of the others' bids. Bidding $b_i = \theta_i$ is therefore optimal regardless of others' bids. Since this is true for each agent, $\ut$ is a Nash equilibrium profile for the indicated valuation profile.\qed

\textbf{Proof of Lemma \ref{lem:sigmasps-lipschitz}:}

Without loss of generality, let $\sigma_S(\ub) \geq \sigma_S(\ub')$. Let $\ua$ and $\ua'$ be the optimal allocations under bid profiles $\ub$ and $\ub'$, respectively. Then
$$
\sum_i \vas(a_i, b'_i) = \sum_i b'_i U(a_i) \stackrel{(\star)}{\leq} \sigma_S(\ub') \leq \sigma_S(\ub) = \sum_i b_i U(a_i),
$$
where ($\star$) follows because $\sigma_S(\ub')$ is the value under the optimal allocation for $\ub'$, and so
\begin{eqnarray*}
  |\sigma_S(\ub)-\sigma_S(\ub')| & \leq & \left|\sum_{\substack{i}} b_i U(a_i) -\sum_{\substack{i}} b'_i U(a_i)\right| = \left|\sum_{\substack{i}} (b_i-b'_i) U(a_i)\right| \\ \nonumber
     & \leq & U(1) \sum_{\substack{i}} |b_i-b'_i| = U(1) \|\ub-\ub'\|_1.
\end{eqnarray*}
The Cauchy-Schwarz inequality gives $\|\ub-\ub' \|_1 \leq \sqrt{n} \|\ub-\ub' \|$, in terms of the Euclidean norm. This proves the Lipschitz property of the mapping $\ub \mapsto \sigma_S(\ub)$ with constant $U(1) \sqrt{n}$.

Define $\sigma_{S, -i}(\ub_{-i}) = \sum_{j: j \ne i} \vas(a_{-i,j}^*, b_j)$. Clearly, $\sigma_{S, -i}$ is also Lipschitz with the same constant $U(1) \sqrt{n}$. For two bid profiles $\ub$ and $\ub'$, applying the definition of $p_{S}$ in (\ref{eqn:p-S}), we get
\begin{eqnarray*}
  |p_{S}(\ub) - p_{S}(\ub')| 
    & \leq & \sum_{i} |\sigma_{S,-i}(\ub_{-i}) - \sigma_{S,-i}(\ub_{-i}')| + (n-1) |\sigma_S(\ub) - \sigma_S(\ub')| \\
    & \leq & U(1)\sqrt{n} \left[\sum_{i} || \ub_{-i} - \ub_{-i}'  || + (n-1)
    || \ub - \ub'  || \right]\\
    & \leq &  (2 U(1) n \sqrt{n}) ~|| \ub - \ub'  ||,
\end{eqnarray*}
where the last inequality follows because $|| \ub_{-i} - \ub'_{-i} || \leq || \ub - \ub' ||$ for all $i$. This proves the Lipschitz property of the mapping $\ub \mapsto p_S(\ub)$ with constant $2 U(1) n \sqrt{n}$.\qed
\newpage
\textbf{Proof of Lemma \ref{lem:simplifyVP}:}

  (1) We show that $ c_0 = c_1 = 0 $ by setting $ \ut = \ue_0 $ and $ \ut = \ue_1 $ in (F) and (VP), where $ \ue_k = (1,1,\ldots,1,0,\ldots,0) $ with $k$ $1$'s. Setting $ \ut = \ue_0 $ in (F), we get $ nc_0 \leq p_S(\ue_0) = 0 $. Setting $ \ut = \ue_0 $ in (VP) yields $ c_0 \geq -v_i(a_i^*)\ \forall i$. When $\ut = \ue_0$, there is no allocation, $a_i^*=0$ for all $i$, and we get $ c_0 \geq -v_{i}(0) = 0$. Hence $c_0 = 0$.

Setting $ \ut = \ue_1 $ in (F), we get $ (n-1) c_1 \leq p_S(\ue_1) = 0$. Setting $\ut = \ue_1$ and $i=2$ in the (VP) constraint, we get $ c_1 \geq q_2(\ue_1) = -v_2(0) = 0 $. Therefore, $c_1 = 0$.

(2) We first show the forward implication. From (\ref{eqn:ni}), we know that $q_k(\ue_{k-1})=0$, since $v_k(0) = 0$ and $\vas(\cdot,0)=0$. Thus, from (VP), we have $r_k(\ue_{k-1}) \geq 0$. Substituting $k = 3,\ldots,n$, we obtain (\ref{eqn:simplify-VP}).

Now we show the reverse implication. Let $\ut \in \hT$ be a Nash equilibrium for the valuation function profile $v_1,\ldots,v_n$. From \citep[Lem. 1]{Guo}, if $\sum_{i=2}^k c_i \geq 0 \ \forall k=2,\ldots,(n-1)$, then $c_2b_2 + \ldots + c_{i-1}b_{i-1} + c_ib_{i+1} + \ldots + c_{n-1}b_n \geq 0 \ \forall \ub \in \hT, \forall i$, i.e., $r_i(\ub_{-i}) \geq 0 \ \forall \ub \in \hT ,\forall i$. Hence $r_i(\ut_{-i}) \geq 0, \forall i$. Now, consider $q_i(\ut)$:
\begin{align*}
-q_i(\ut) &= v_i(a_i^*(\theta_i,\ut_{-i})) + \sum_{j \ne i}\vas(a_j^*(\theta_i,\ut_{-i}),\theta_j)-\sum_{j \ne i}\vas(a_{-i,j}^*(\ut_{-i}),\theta_j) \\
&\geq v_i(a_i^*(0,\ut_{-i})) + \sum_{j \ne i}\vas(a_j^*(0,\ut_{-i}),\theta_j) -\sum_{j \ne i}\vas(a_{-i,j}^*(\ut_{-i}),\theta_j),
\end{align*}
where the inequality holds because $\theta_i$ is the best response against $\ut_{-i}$ (Nash equilibrium); replacing $\theta_i$ by $0$ should yield a lower value. Now, the right-hand side of the inequality is zero because: (1) assumption $\vas(a,0)=0$ implies that $a_i^{*}(0,\ut_{-i})=0$, and $v_i(0) = 0$ implies that the first term $v_i(a_i^*(0,\ut_{-i})) = 0$, and (2) $\theta_i = 0$ implies that presence of agent $i$ does not affect the outcome for the other agents, thereby resulting in the second and third terms canceling each other. Therefore, we have $q_i(\ut) \leq 0$. Since, we already know $r_i(\ut_{-i}) \geq 0$, we have $r_i(\ut_{-i}) \geq q_i(\ut)$.\qed

\begin{rmk}
We put the bids $\ue_k, k = 0, 1, \ldots, n$ to good use in the proof above. Though some of these bids may not be realizable -- for example, $\ue_0$ and $\ue_1$ are not realizable because our Assumption \ref{ass_efficient_NE}(a) forces at least two nonzero bidders in any (efficient) Nash equilibrium -- we may still use these bids because they belong to the closure of the set of Nash equilibria. These bids may be approached through a sequence of realizable bids.
\end{rmk}

\begin{rmk}
Setting $\uc=0$, we see that Clarke's pivotal payments satisfy (VP) even in the SSVCG setting with surrogate valuations, where $v_i$'s are not fully known to the social planner.
\end{rmk}
\newpage
\textbf{Addendum to Section \ref{subsec:surrogate}:}
\begin{lem}
 Consider $\vas(a,\theta)=\theta a^{1-\alpha}$ for some $\alpha\in(0,1)$. Then the expression $\frac{p_S(\ut)}{\sigma_S(\ut)}$ is maximized at $\ut=\mu\ue_n$ for any $\mu>0$.
\end{lem}
\begin{pf}
Recall that the efficient allocation solves (\ref{Soc_plan_prob_a}). From the KKT conditions, the efficient allocation is $a_i=\theta_i^{1/\alpha}/\sum_{j=1}^n\theta_j^{1/\alpha}$, $i=1,2,\ldots,n$. Substituting $a_i$ in $\sigma_s(\ut)$, we get $\sigma_S(\ut)=\|\ut\|_{\frac{1}{\alpha}}$. Here, $\|.\|_p$ denotes the $L_p$-norm. Thus we have $p_S(\ut)/\sigma_S(\ut)=\sum_{j=1}^n\|\ut_{-j}\|_{\frac{1}{\alpha}}/\|\ut\|_{\frac{1}{\alpha}}-(n-1)$. To find the maximizer, we differentiate this expression and equate it to zero, to get
\begin{equation}\label{eqn:max-ps/sigmas}
 \frac{\partial}{\partial\theta_i}\left(\frac{p_S(\ut)}{\sigma_S(\ut)}\right)=\frac{\theta_i^{\frac{1-\alpha}{\alpha}}}{\|\ut\|^2_{\frac{1}{\alpha}}}\left(\sum_{j\ne i}\frac{\|\ut\|_{\frac{1}{\alpha}}}{\|\ut_{-j}\|^{\frac{1-\alpha}{\alpha}}_{\frac{1}{\alpha}}}-\sum_{j=1}^n\frac{\|\ut_{-j}\|_{\frac{1}{\alpha}}}{\|\ut\|^{\frac{1-\alpha}{\alpha}}_{\frac{1}{\alpha}}}\right)=0.
\end{equation}
(\ref{eqn:max-ps/sigmas}) is satisfied for all $i$ if and only if $\ut=\mu\ue_n$. This can be observed as follows:
\begin{itemize}
 \item (\ref{eqn:max-ps/sigmas}) holds for all $i$ if and only if the following quantity is a constant for all $i$.
$$
  \sum_{j=1}^n\frac{\|\ut\|_{\frac{1}{\alpha}}}{\|\ut_{-j}\|^{\frac{1-\alpha}{\alpha}}_{\frac{1}{\alpha}}}-\frac{\|\ut\|_{\frac{1}{\alpha}}}{\|\ut_{-i}\|^{\frac{1-\alpha}{\alpha}}_{\frac{1}{\alpha}}}-\sum_{j=1}^n\frac{\|\ut_{-j}\|_{\frac{1}{\alpha}}}{\|\ut\|^{\frac{1-\alpha}{\alpha}}_{\frac{1}{\alpha}}}.
$$
 \item This occurs if and only if $\frac{\|\ut\|_{\frac{1}{\alpha}}}{\|\ut_{-i}\|^{\frac{1-\alpha}{\alpha}}_{\frac{1}{\alpha}}}$ is a constant for all $i$.
 \item This in turn occurs if and only if $\|\ut_{-i}\|_{\frac{1}{\alpha}}$ is a constant for all $i$, which in turn occurs if and only if $\theta_i$ is a constant for all $i$.
\end{itemize}
Thus (\ref{eqn:max-ps/sigmas}) is satisfied for all $i$ only when $\theta_1=\ldots=\theta_n=\mu$.

We now use second order condition to show that $\ut=\mu\ue_n$ is indeed the maximizer. The entries of the Hessian matrix $\nabla^2(p_S(\ut)/\sigma_S(\ut))$ at $\ut=\mu\ue_n$ can be shown to be
$$
  \left(\nabla^2(p_S(\mu\ue_n)/\sigma_S(\mu\ue_n))\right)_{ij}=\begin{cases}\frac{1}{\mu^2}\frac{1-1/\alpha}{n^{1+\alpha}(n-1)^{1-\alpha}}&i=j\\\frac{1}{\mu^2}\frac{1/\alpha-1}{n^{1+\alpha}(n-1)^{2-\alpha}}&i\ne j.\end{cases}
$$
The term $\ut^T\left[\nabla^2(p_S(\mu\ue_n)/\sigma_S(\mu\ue_n))\right]\ut$ equals $0$ for every $\ut=\{\nu\ue_n,\nu\in\mathbb{R}\}$, and is negative otherwise. This indicates that the function $f(\ut)=p_S(\ut)/\sigma(\ut)$ is strictly concave in every line passing through $\mu\ue_n$, except on the line $\ut=\{\nu\ue_n,\nu\in\mathbb{R}\}$, where it is a constant. Thus $\ut=\mu\ue_n$ maximizes $p_S(\ut)/\sigma_s(\ut)$.\qed
\end{pf}

The worst-case for SSVCG thus occurs at $\ut=\mu\ue_n$. We now bound the objective.
\begin{prop}
 $\frac{p_S(\ue_n)}{\sigma_S(\ue_n)}\uparrow 1-\alpha$ when $n\rightarrow\infty$.
\end{prop}
\begin{pf}
\begin{multline}\label{eqn:ps/sigmas}
 \frac{p_S(\ue_n)}{\sigma_S(\ue_n)}=n\left(1-\frac{1}{n}\right)^\alpha-(n-1)=n\left(1-\frac{\alpha}{n}-\frac{\alpha(1-\alpha)}{(2!)n^2}-...\right)-(n-1)\\=(1-\alpha)-\left(\frac{\alpha(1-\alpha)}{(2!)n}+\frac{\alpha(1-\alpha)(2-\alpha)}{(3!)n^2}+...\right)
\end{multline} 
where the second equality occurs by binomial expansion. So $p_S(\ue_n)/\sigma_S(\ue_n)$ increases in $n$. It increases to $1-\alpha$ when $n\rightarrow\infty$.\qed
\end{pf}
This result gives an impression that the use of linear rebates is unwarranted to minimize the worst-case objective, because our choice of the worst-case objective can indeed be driven as close to $0$ as we wish, just by choosing $\vas(a,\theta)=\theta a^{1-\alpha}$ with $\alpha$ chosen close to $1$. We now show that the reduction in the objective function value occurs because of an increase in $\sigma_S(\ut)$ at the Nash equilibrium point, and not due to the reduced surplus.

\begin{prop}\label{prop:sigmas-increase-NE}
 Consider $\ut_{NE}=(\mu_{NE})\ue_n$ to be the Nash equilibrium point. Then,
\begin{enumerate}
 \item $p_S(\ut_{NE})\uparrow v'(1/n)$ as $n\rightarrow\infty$.
 \item $\sigma_S(\ut_{NE})=\frac{v'(1/n)}{1-\alpha}$.
\end{enumerate}
\end{prop}
\begin{pf}
The Nash equilibrium occurs at the $\ut$ satisfying $v_i'(a_i)=\vas'(a_i,\theta_i)$ for all $i$ \citep[Cor. 1]{Johari09}. In our case, we have $\vas(a,\theta)=\theta a^{1-\alpha}$. Thus the Nash equilibrium occurs at $\ut$ where $v_i'(a_i)=(1-\alpha)\theta_i/a_i^\alpha$ holds for all $i$. So we have $(\ut_{NE})_i=\frac{v_i'(a_i)}{1-\alpha}a_i^\alpha$. When $\ut=\ut_{NE}=(\mu_{NE})\ue_n$ is a Nash equilibrium point, we have $a_i=1/n$ for all $i$, and thus $v_i'(1/n)$ must be equal for all $i$. So $\mu_{NE}=\frac{v'(1/n)}{(1-\alpha)n^\alpha}$.

We compute $p_S(\ut_{NE})$ as
$$
  p_S(\ut_{NE})=(\mu_{NE})n^\alpha\left(n\left(1-\frac{1}{n}\right)^\alpha-(n-1)\right)\leq(\mu_{NE})n^\alpha(1-\alpha)=v'(1/n)
$$
where the inequality occurs by tracing the same steps in (\ref{eqn:ps/sigmas}). The same steps point that $p_S(\ut_{NE})\uparrow v'(1/n)$ when $n\rightarrow\infty$.

Now we compute $\sigma_S(\ut_{NE})$ as
$$
  \sigma_S(\ut_{NE})=\mu_{NE}\sum_{i=1}^n(1/n)^{1-\alpha}=nv'(1/n)\frac{(1/n)^{1-\alpha}}{(1-\alpha)n^\alpha}=\frac{v'(1/n)}{1-\alpha}.
$$
Hence the result.\qed
\end{pf}

Proposition \ref{prop:sigmas-increase-NE} thus shows that the surplus at $\ut_{NE}$ approaches a constant, as $n\rightarrow\infty$, independent of $\alpha$, but $\sigma_S(\ut_{NE})$ approaches $p_S(\ut_{NE})/(1-\alpha)$. This explains why we scaled our worst-case objective by $1/(1-\alpha)$ to get the normalized plots in Figures \ref{sqrt_figure} and \ref{consolidate_figure}. Varying $\alpha$ changes $\sigma_S(\ut)$, but the surplus is quite stable across $\alpha$. Thankfully, linear rebates return a significant fraction of the surplus.

\end{document}